\documentclass[aps,prd,twocolumn,floatfix,noshowpacs,tightenlines,noshowkeys,superscriptaddress,amsmath,amssymb,nofootinbib]{revtex4}
\usepackage{amssymb,amsbsy,epsfig,color,graphicx}
\usepackage{color}
\usepackage{[longtable}
\usepackage{array}
\usepackage{dcolumn}   
\usepackage{cellspace}
\usepackage{mathtools}
\usepackage{amstext}
\usepackage{amssymb}
\usepackage{stmaryrd}
\usepackage{stackrel}
\usepackage{graphicx}
\usepackage{esint}
\usepackage[utf8]{inputenc}
\usepackage{blindtext}
\usepackage{float}
\restylefloat{table}
\usepackage{booktabs}
\usepackage{enumitem}
\usepackage{bbold}
\usepackage{slashed}
\usepackage{hyperref}
\usepackage{etoolbox} 
\usepackage{lipsum} 
\usepackage[capitalize]{cleveref}
\usepackage{multirow}
\usepackage[caption=false]{subfig}
\allowdisplaybreaks

\def\be{\begin{equation}}
\def\ee{\end{equation}}
\def\bea{\begin{eqnarray}}
\def\eea{\end{eqnarray}}

\newcommand{\ba}{\begin{eqnarray}}
\newcommand{\ea}{\end{eqnarray}}

\renewcommand\[{\begin{equation}}
\renewcommand\]{\end{equation}}
\makeatletter

\appto{\appendix}{%
\@ifstar{\def\theequation@prefix{A.}}%
{}%
}
\makeatother

\hypersetup{pdfstartview=FitV,colorlinks=true,linkcolor=midblue,citecolor=midblue,filecolor=midblue,urlcolor=midblue}

\definecolor{midblue}{rgb}{0,0,0.5}

\begin{document}

\title{Quantum interference in external gravitational fields beyond General Relativity}

\author{Luca Buoninfante}
\email{buoninfante.l.aa@m.titech.ac.jp}
\affiliation{Department of Physics, Tokyo Institute of Technology, Tokyo 152-8551, Japan}

\author{Gaetano Lambiase}
\email{lambiase@sa.infn.it}
\affiliation{INFN - Sezione di Napoli, Gruppo collegato di Salerno, I-84084 Fisciano (SA), Italy}
\affiliation{Dipartimento di Fisica ``E.R. Caianiello'', Universit\`a degli Studi di Salerno, I-84084 Fisciano (SA), Italy}

\author{Luciano Petruzziello}
\email{lupetruzziello@unisa.it}
\affiliation{INFN - Sezione di Napoli, Gruppo collegato di Salerno, I-84084 Fisciano (SA), Italy}
\affiliation{Dipartimento di Ingegneria Industriale, Universit\`a degli Studi di Salerno, Fisciano (SA) 84084, Italy}


\begin{abstract}
In this paper, we study the phenomenon of quantum interference in the presence of external gravitational fields described by alternative theories of gravity. We analyze both non-relativistic and relativistic effects induced by the underlying curved background on a superposed quantum system. In the non-relativistic regime, it is possible to come across a gravitational counterpart of the Bohm-Aharonov effect, which results in a phase shift proportional to the derivative of the modified Newtonian potential. On the other hand, beyond the Newtonian approximation, the relativistic nature of gravity plays a crucial r\^ole. Indeed, the existence of a gravitational time dilation between the two arms of the interferometer causes a loss of coherence that is in principle observable in quantum interference patterns. We work in the context of generalized quadratic theories of gravity to compare their physical predictions with the analogous outcomes in general relativity. In so doing, we show that the decoherence rate strongly depends on the gravitational model under investigation, which means that this approach turns out to be a promising test bench to probe and discriminate among all the extensions of Einstein's theory in future experiments.
\end{abstract}

\maketitle


\section{Introduction}

Einstein's General Relativity (GR) has gone through many challenges in the last century, but it has always been confirmed by high-precision experiments which have verified many of its predictions~\cite{-C.-M.}. The recent observation of gravitational waves from binary merger represents one of the most astonishing of its achievements~\cite{Abbott:2016blz}.

Despite its great success, there are conceptual problems which have not found a definite answer yet. For instance, by focusing on galactic and cosmological scales, self-consistent and complete descriptions for dark matter and dark energy (which are both compatible with experimental data) are still missing. Furthermore, in the short-distance (ultraviolet) regime, GR turns out to be classically incomplete due to the presence of cosmological and black hole singularities, whereas from a quantum point of view it is a non-renormalizable theory, which thus lacks predictability at high energies. From an experimental point of view, what we can say is that our knowledge about short-distance gravity is extremely limited; indeed, Newton's law has been tested only up to micrometer scales~\cite{Kapner:2006si} and the smallest masses for which the gravitational coupling has been measured are of the order of $100$ milligrams~\cite{Westphal:2020okx}.

In the past years, these fundamental open issues have channeled a huge amount of efforts towards the quest for a consistent ultraviolet completion of GR. One of the most straightforward approaches consists in generalizing the Einstein-Hilbert action by including terms which are quadratic in the curvature invariants, i.e. $\mathcal{R}^2,$ $\mathcal{R}_{\mu\nu}\mathcal{R}^{\mu\nu}$ and $\mathcal{R}_{\mu\nu\rho\sigma}\mathcal{R}^{\mu\nu\rho\sigma}.$ The first remarkable achievements in the framework of quadratic gravity date back to 1977 with the results obtained by Stelle~\cite{-K.-S.}, who proved that a gravitational theory described by the Einstein-Hilbert action with the addition of the terms $\mathcal{R}^2$ and $\mathcal{R}_{\mu\nu}\mathcal{R}^{\mu\nu}$ is power-counting renormalizable. At the same time, however, such a gravitational model hides undesirable features, such as the emergence of a massive spin-$2$ ghost degree of freedom that violates unitarity (when standard quantization prescriptions are implemented~\cite{Anselmi}). Despite the presence of the ghost field, the above theory can be regarded as an effective field theory valid at the energy scales below the cut-off represented by the mass of the ghost. Another important accomplishment in the framework of quadratic gravity is given by the Starobinski model of inflation~\cite{starobinski}, which is in good agreement with the current data, even though in this case the only quadratic part of the action is $\mathcal{R}^2$. In addition to that, it is worth observing that gravitational actions with quadratic curvature corrections were recently considered also in other different scenarios~\cite{capoz1,lamb1,lamb2,Buoninfante:2019uwo,Lambiase:2020vul,Asorey:1996hz,lv,Lambiase:2016bjy,Buoninfante:2020qud,neut,wagner}.

The results mentioned so far were obtained for {\it local} quadratic theories of gravity, whose corresponding Lagrangians depend polynomially on the derivative operator. Recently, also {\it nonlocal} quadratic modifications have burst into the spotlight, as the presence of nonlocal (i.e. non-polynomial) form factors in the gravitational action can help both to solve the problem of ghosts and to improve the ultraviolet behavior of the quantized theory. For this vast topic, we remand the interested reader to Refs.~\cite{Krasnikov,Kuzmin,Tomboulis:1997gg,Biswas:2005qr,Modesto:2011kw,Biswas:2011ar,Biswas:2013cha,Biswas:2016etb,Edholm:2016hbt,Koshelev:2017tvv,Koshelev:2020foq,Buoninfante:2018rlq,Buoninfante:2018stt,Buoninfante:2018xif,Buoninfante:2018xiw,Frolov:2015bta,Boos:2018bxf,Kolar:2020bpo,Dengiz:2020xbu,Boos:2021suz}.

In this paper, our aim is to investigate the differences between GR and several extended theories of gravity by resorting to the phenomenon of \textit{quantum interference}. Specifically, we will consider both non-relativistic and relativistic effects induced by a modified gravitational model onto a quantum interference experiment to compare the results with the case of Eistein's theory. The interplay between GR and quantum interference has already been addressed, both from a theoretical and a phenomenological perspective. As a matter of fact, in the non-relativistic regime (Newtonian approximation), the relevant effect is a Bohm-Aharonov-like phase shift that is proportional to the derivative of the gravitational potential, as discovered for the first time in 1975 in a laboratory test devised by Colella, Overhauser and Werner, better known under the name of COW experiment~\cite{Colella:1975dq}. 
On the other hand, in the relativistic domain the existence of time dilation entails more drastic effects on a quantum superposition. Indeed, in Refs.~\cite{Zych:2011hu,Pikovski:2013qwa,Zych-thesis} it was shown that a (gravitational) time dilation between the two arms of an interferometer can cause a loss of coherence in the interference pattern, thereby giving rise to \textit{decoherence}. This quantum manifestation has not been directly observed up to now, since a detectable loss of coherence would require either a large travel-time or a large distance between the two arms of the interferometer (see Sec.~\ref{quant-int-beyond-sec}). However, a decoherence mechanism originated by time dilation was recently found out in a similar experiment~\cite{Margalit}, where a Stern-Gerlach interferometer under the influence of an external inhomogeneous magnetic field was employed to simulate the effect of time dilation on the spin precession of the examined system (atom chip). 

To comply with the aforementioned purposes, the paper is organized as follows: in Sec.~\ref{mach-zeh-sec} we analyze the quantum mechanical setup, focusing in particular on the physics behind a Mach-Zehnder interferometer and on the complementary concepts of interferometric visibility and which-way information. In Sec.~\ref{massive clock-sec} we adapt the above setting to a configuration in which the interferometer is embedded in a (classical) weak and static gravitational field. Then, we derive the Hamiltonian of a quantum system in curved backgrounds and rely on a two-level system as a simple realization of a quantum massive ``clock''. Equipped with this knowledge, we manage to discuss both COW and gravitational time dilation effects associated with a generic, linearized and static spacetime metric. Section~\ref{quant-int-beyond-sec} is devoted to the introduction of several extended theories of gravity; for each of them, we exhibit the corresponding modified Newtonian potential which has to be exploited for the computation of detection probabilities and interferometric visibility. In Sec.~\ref{discuss-sec}, we compare the new predictions related to alternative theories of gravity with the current experimental data. In this respect, we point out that the decoherence rate triggered by time dilation strongly depends on the gravitational model under investigation. Furthermore, we comment on the fact that such an intriguing aspect can provide a valuable test bench to probe and discriminate among several alternative theories in future experiments. 
Finally, Section~\ref{concl-sec} contains concluding remarks and outlook. In Appendix~\ref{schr-curved-sec}, we allocate the mathematical details of the derivation of the Hamiltonian for a quantum system in an external, linearized and static spacetime metric, whilst in Appendix~\ref{gen-poisson-sec} we briefly review the linearized (weak-field) limit of generalized quadratic gravitational theories and their modified Newtonian potentials.
 
Before ending this preliminary Section, let us stress that in this paper we only work with \textit{external} and \textit{classical} gravitational fields (namely, we study quantum systems on \textit{classical} backgrounds). In other terms, we follow a semi-classical approach and reasonably assume that the self-gravity of a given quantum system is negligible in the analyzed physical setting. On the other hand, it must be said that several articles recently appeared in literature have accounted for self-gravity effects, both at the classical~\cite{Bahrami:2014gwa,Buoninfante:2017kgj,Buoninfante:2017rbw} and quantum level~\cite{Bose:2017nin,Marletto:2017kzi,Belenchia:2018szb,Marshman:2019sne}.

Throughout this work, we adopt the positive convention for the metric signature, that is $\eta={\rm diag}(-1,+1,+1,+1).$

\section{Quantum complementarity}\label{mach-zeh-sec}

In quantum mechanics, there are physical properties that cannot be simultaneously accessed 
with arbitrary precision; these quantity are addressed as \textit{complementary}, and mathematically they correspond to non-commuting operators. One particular example of complementarity is realized when observing interference \textit{versus} the availability of which-way information: these notions are mutually exclusive in any interference experiment, as the double-slit or the Mach-Zehnder setup. 

For later convenience, we need to briefly review the main characteristics of a Mach-Zehnder interferometer; in so doing, we closely follow Refs.~\cite{Englert:1996zz,Zych-thesis}. 


\subsection{Mach-Zehnder interferometer}

\begin{figure}[t]
	\includegraphics[scale=0.225]{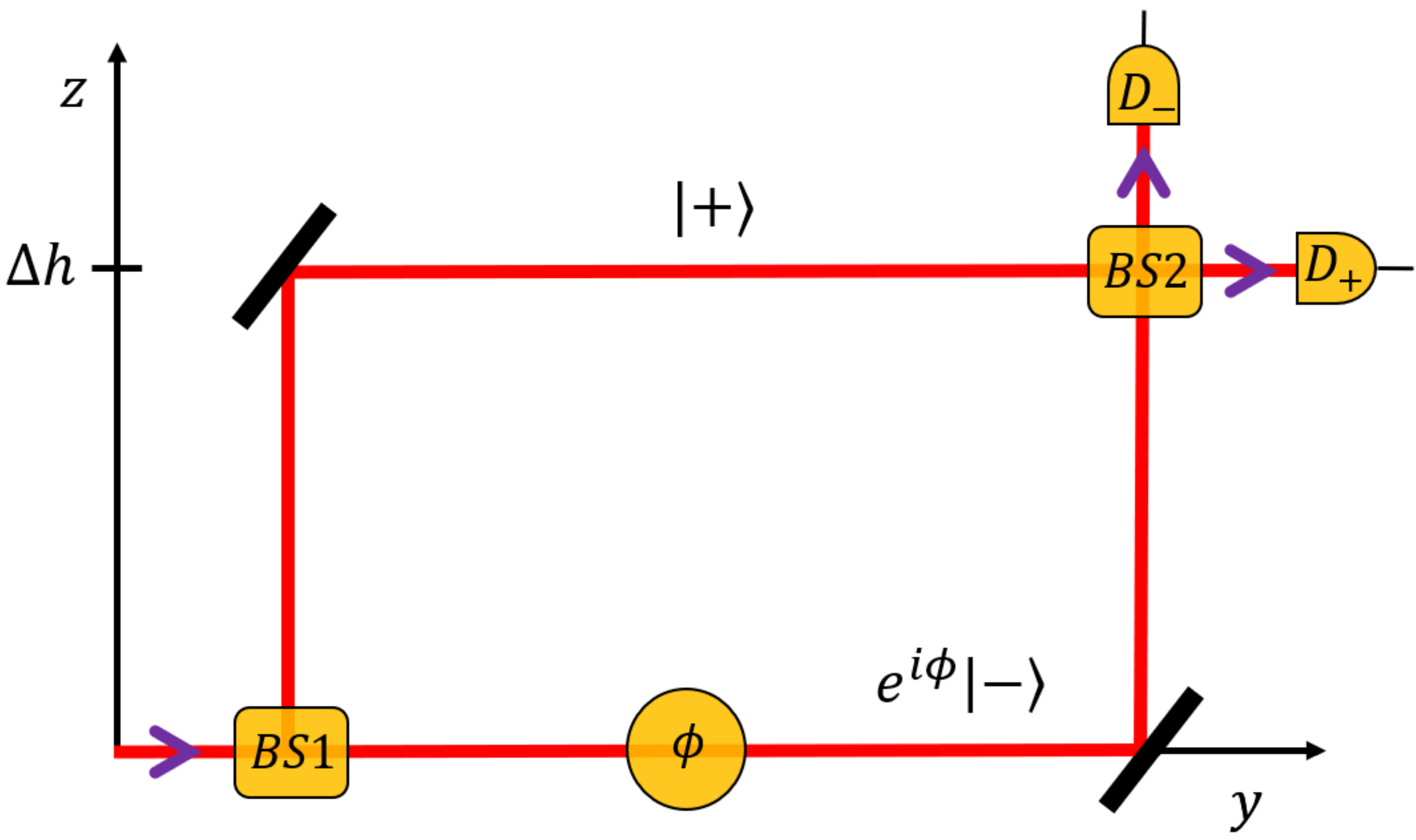}
	\centering
	\protect\caption{This figure illustrates a two-way interferometer known as Mach-Zehnder setup. It is made up of two beam splitters BS1, BS2 and two detectors $D_{\pm}$. The superposed system travels along the two paths that are labeled by $\left| \pm \right\rangle,$ and it can generally acquire a relative (controllable) phase shift $\phi$. The two detectors $D_{\pm}$ measure the degree of interference which appears on the screen as an interference pattern.}\label{fig1}
\end{figure}

Let us consider a two-dimensional Hilbert space $\mathcal{H}_1$ with an orthonormal basis $\left\lbrace \left| +\right\rangle, \left| -\right\rangle \right\rbrace.$ Such states can be used to describe a superposed quantum system traveling along the two arms of an interferometer, as the Mach-Zehnder setup shown in Fig.~\ref{fig1}.

The two detectors $D_{\pm}$ in Fig.~\ref{fig1} quantify the degree of interference by measuring the two physical observables $\sigma_x$ and $\sigma_z$ ($x$ and $z$ Pauli matrix respectively). The former is defined as
\begin{eqnarray}
\sigma_x:=\left| + \right\rangle \left\langle - \right| +\left| - \right\rangle \left\langle + \right|\,,
\end{eqnarray}
and we can say that it is measured
when the detection of the interfering system in $D_{\pm}$ is associated with an outcome $\pm1$. Indeed, one can easily prove that the available output states $\left| + \right\rangle \pm \left| - \right\rangle$ are eigenstates of $\sigma_x$ with eigenvalues $\pm1,$ namely $\sigma_x  (\left| + \right\rangle \pm \left| - \right\rangle)=\pm (\left| + \right\rangle \pm \left| - \right\rangle).$  

On the other hand, the second observable quantifies the which-path information and it is given by
\begin{eqnarray}
\sigma_z:=\left| + \right\rangle \left\langle + \right| -\left| - \right\rangle \left\langle - \right|\,.
\end{eqnarray}
Such quantity has eigenstates $\left| \pm \right\rangle$ and respective eigenvalues $\pm1,$ which means $\sigma_z\left| \pm \right\rangle=\pm \left| \pm \right\rangle.$

Therefore, the observables $\sigma_x$ and $\sigma_z$ are two anti-commuting
operators that take into account the quantum complementarity between interference
and which-way information.
To move forward, we still have to tackle an important aspect, that is reflected into the main differences between quantum interferometry for pure and mixed states, as well as interferometry for single and composite systems.

\subsection{Pure states}

If the interfering system is in a \textit{pure} state, we can write
\begin{eqnarray}
\left| \psi \right\rangle=a \left| + \right\rangle +e^{i\phi}b\left| - \right\rangle\,,\label{pure-state}
\end{eqnarray}
where $a,b\in \mathbb{R}$ satisfy $a^2+b^2=1,$ whilst $\phi\in [0,2\pi)$ is a controllable phase shift between the two different paths. The probabilities of obtaining the two values $\pm$ when
measuring $\sigma_x$ are
\begin{eqnarray}
P_{\sigma_x=\pm1}(\phi)=\left|\left\langle  \psi \right| \left(\frac{\left| + \right\rangle \pm \left| - \right\rangle}{\sqrt{2}}\right) \right|^2=\frac{1}{2}\pm a\,b\cos \phi\,.
\end{eqnarray}
For the observable $\sigma_z$, instead, one has
\begin{eqnarray}
P_{\sigma_z=+1}=\left|\left\langle  \psi \right|  \left.+ \right\rangle \right|^2=a^2\,,&&\\ P_{\sigma_z=-1}=\left|\left\langle  \psi \right| \left. - \right\rangle \right|^2=b^2\,.
\end{eqnarray}
Now, we can introduce two fundamental quantities that turn out to be convenient to properly describe quantum complementarity. 
The \textit{interferometric visibility} (or simply \textit{visibility}) is defined as
\begin{eqnarray}
\mathcal{V}:=\frac{{\rm max}_\phi P_{\sigma_x=\pm}-{\rm min}_\phi P_{\sigma_x=\pm1}}{{\rm max}_\phi P_{\sigma_x=\pm}+{\rm min}_\phi P_{\sigma_x=\pm1}}\,.\label{def-visibility}
\end{eqnarray}
Since ${\rm max}_\phi \cos\phi=1$ and ${\rm min}_\phi \cos\phi=-1,$ for pure states the interferometric visibility reads
\begin{eqnarray}
\mathcal{V}=2ab\,.\label{visib-pure}
\end{eqnarray}
Furthermore, the \textit{predictability} of a measurement for the observable $\sigma_z$ is identified with
\begin{eqnarray}
\mathcal{P}:=\left| P_{\sigma_z=+1}-P_{\sigma_z=-1} \right|=|a^2-b^2|  \,.\label{predict-pure}
\end{eqnarray}
The pure state~\eqref{pure-state} is normalized; hence, it is straightforward to observe that 
\begin{eqnarray}
\mathcal{V}^2+\mathcal{P}^2=1 \,.\label{P+V=1}
\end{eqnarray}
The physical meaning of this last equation lies in the fact that a non-zero predictability of the two paths necessarily implies a non-maximal interferometric visibility of the interference pattern and vice-versa. In other words, a non-maximal predictability entails the appearance of a visible interference pattern.

\subsection{Mixed states}

In the case of \textit{mixed} states, Eq.~\eqref{P+V=1} is generalized to
\begin{eqnarray}
\mathcal{V}^2+\mathcal{P}^2\leq 1 \,\label{P+V<1}\,,
\end{eqnarray}
consistently with the interpretation according to which mixed states are characterized by an incomplete knowledge about the physical state. 
In order to reach Eq.~\eqref{P+V<1}, we have to rely on the density matrix formalism. Starting from
\begin{eqnarray}
\rho&=&a^2 \left| + \right\rangle \left\langle + \right|
+b^2\left| - \right\rangle \left\langle - \right|\nonumber \\[2mm]
&+&c^*e^{-i\phi}\left| + \right\rangle \left\langle - \right|
+ce^{i\phi}\left| - \right\rangle \left\langle + \right| \,,\label{dens-matrix}
\end{eqnarray}
we compute $\rho^2$ and impose\footnote{Recall that the trace operation is given by ${\rm Tr}\left\lbrace \rho^2 \right\rbrace = \left\langle+|\rho^2|+ \right\rangle + \left\langle-|\rho^2|- \right\rangle$ and it defines the purity of a state. The inequality ${\rm Tr}\left\lbrace \rho^2 \right\rbrace\leq 1 $ holds true for any mixed state, and it is saturated only for pure states that indeed satisfy ${\rm Tr}\left\lbrace \rho^2 \right\rbrace =1.$} ${\rm Tr}\left\lbrace \rho^2\right\rbrace \leq 1,$ which yields
\begin{eqnarray}
|c|^2\leq ab\,.\label{trace<1}
\end{eqnarray}
The probabilities to measure $\left| \pm \right\rangle $ for the observable $\sigma_z$ do not depend on the off-diagonal elements, thus leaving the expression for the predictability~\eqref{predict-pure} untouched. Differently from the above picture, the probabilities $P_{\sigma_x=\pm1}$ to detect an interference pattern for the mixed state outlined by the density matrix $\rho$ is given by
\begin{eqnarray}
 P_{\sigma_x=\pm1}&=&\left(\frac{\left\langle +\right|\pm \left\langle -\right| }{\sqrt{2}}\right)\rho\left(  \frac{\left| + \right\rangle\pm \left| - \right\rangle}{\sqrt{2}} \right)\nonumber\\
 & = & \frac{1}{2}\pm |c|\cos(\phi+\alpha)   \,,\label{V-mixed}
\end{eqnarray}
where we have used $c=|c|e^{i\alpha}.$ From the last equation and Eq.~\eqref{def-visibility}, we can deduce the interferometric visibility for a mixed state:
\begin{eqnarray}
\mathcal{V}=2|c|\,.\label{visib-mixed}
\end{eqnarray}
In general, the interferometric visibility coincides with twice the amplitude of the off-diagonal component of the density matrix. As a matter of fact, for a pure state $|c|=ab,$ which recovers Eq.~\eqref{visib-pure}.

Hence, by putting the relations~\eqref{predict-pure} and~\eqref{visib-mixed} together, we obtain the inequality $\mathcal{V}^2+\mathcal{P}^2\leq 1,$ as claimed in Eq.~\eqref{P+V<1}.

\begin{figure}[t]
	\includegraphics[scale=0.225]{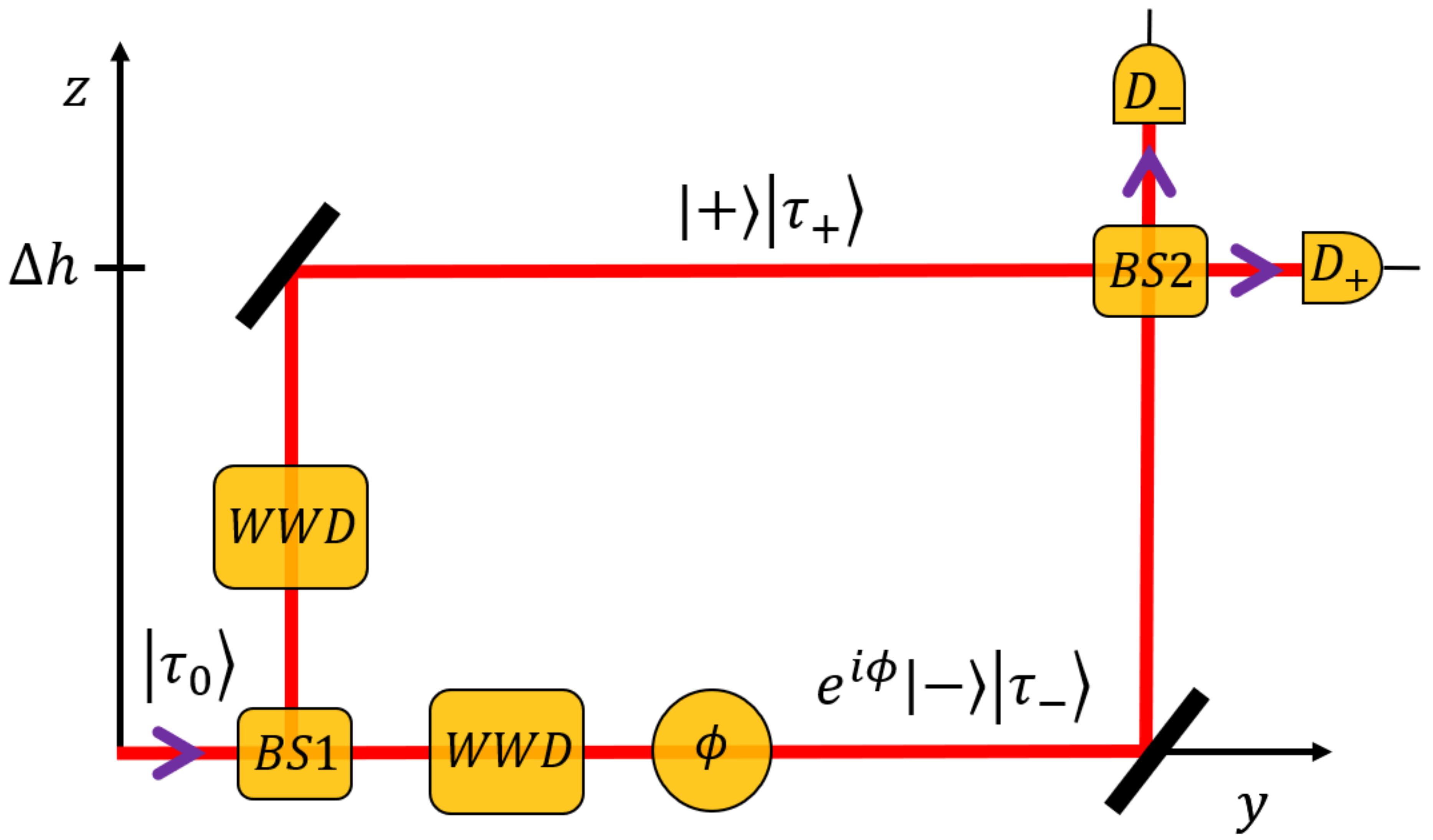}
	\centering
	\protect\caption{This figure illustrates the Mach-Zehnder interferometer in Fig.~\ref{fig1} with the addition of an extra degree of freedom that is able to encode the which-way information. Apart from the two beam splitters BS1, BS2 and the two detectors $D_{\pm},$ we now have a \textit{which-way detector} indicated by WWD. The main difference with respect to the simpler configuration of Fig.~\ref{fig1} lies in the possibility that the interfering system and WWD can become entangled, thereby allowing to obtain the which-way information from WWD. 
		At the same time, the interferometric visibility $\mathcal{V}$ is reduced, which tells us that the complementarity principle still holds even for a composite  system.}\label{fig2}
\end{figure}

\subsection{Composite systems}\label{composite-sec}

Quantum complementarity requires that precise measurements of $\sigma_{x}$ and $\sigma_z$ cannot be performed simultaneously on the same system, as they correspond to two anti-commuting operators. However, so far we have only studied a system made up of a single degree of freedom. We now want to demonstrate that quantum complementarity also applies to systems made up of multiple degrees of freedom. To this aim, we must add a \textit{second} system in our interferometric setup that is called \textit{which-way detector} (WWD) (see Fig.~\ref{fig2} for this configuration). Practically speaking, WWD can be taken as a two-level system.

The Hilbert space of the composite system is labeled as $\mathcal{H} = \mathcal{H}_1\otimes \mathcal{H}_2$, where $\mathcal{H}_1$ is the Hilbert space of the interfering system and $\mathcal{H}_2$ is the Hilbert space of WWD. At the initial reference time $t=0,$ the which-way detector is in some state $\left| \tau_0\right\rangle $ uncorrelated with the interfering system, which implies that the starting state is a product state. The crucial point is that, for $t>0,$ WWD becomes excited and performs a transition to one of the two normalized states $\left| \tau_{\pm}\right\rangle$  depending on the arm traveled by the system. Consequently, the total state of the system becomes \textit{entangled} and reads 
\begin{eqnarray}
\left| \psi \right\rangle=a \left| + \right\rangle\left| \tau_+ \right\rangle +e^{i\phi}b\left| - \right\rangle\left| \tau_-\right\rangle\,.\label{entang-state}
\end{eqnarray}
Since the (composite) state is now entangled, we can measure the two observables $\sigma_x$ and $\sigma_z$ with absolute precision simultaneously. Indeed, we can summarize the measurements in two steps:
\begin{itemize}
	
	\item we access the which-way information by measuring the observable $\mathbb{1}_1\otimes \sigma_z$ on WWD;
	
	\item we detect the interfering system at $D_\pm$ to measure the observable $\sigma_x\otimes \mathbb{1}_2.$

\end{itemize}
In the previous steps, $\mathbb{1}_1$ and $\mathbb{1}_2$ are the two identity operators acting on the states of the Hilbert spaces $\mathcal{H}_1$ and   $\mathcal{H}_2,$ respectively.

At this point, we need to determine the probabilities $P_{\sigma_z=\pm1}$ and $P_{\sigma_x=\pm1}$ for the case of the composite system described by the entangled state~\eqref{entang-state}.
Given the total density matrix  $\rho_\psi= \left| \psi\right\rangle \left\langle \psi \right|,$ one can obtain the density matrix for one of the two subsystems by partial tracing with respect to the other, i.e. $\rho_i = {\rm Tr}_j \left\lbrace \rho_\psi\right\rbrace $ with $i, j = 1, 2,$ thereby remaining with a mixed state.
It is worth stressing that the states $\left| \tau_+\right\rangle $, $\left| \tau_-\right\rangle $ are in general \textit{not} orthogonal (i.e. $\left\langle \tau_+\right.\left| \tau_-\right\rangle\neq 0$); thence, it comes in handy to consider an orthogonal basis in $\mathcal{H}_2$ such that
\begin{eqnarray}
&&\left|\tau_\pm \right\rangle= A_\pm\left|u \right\rangle+B_\pm\left|v\right\rangle\,,\nonumber\\[2mm]
&&\left\langle u\right.\left| v\right\rangle=\delta_{uv}\,,\quad |A_\pm|^2+|B_\pm|^2=1\,. \label{orth-basis-H_2}
\end{eqnarray}
By using the above orthonormal basis, we can readily trace over the Hilbert space $\mathcal{H}_2$ and show that the reduced density matrix $\rho_1$ for the interfering system is
\begin{eqnarray}
\!\!\!\!\!\!\!&&\rho_1=a^2 \left| + \right\rangle \left\langle + \right|
+b^2 \left| - \right\rangle \left\langle - \right| \nonumber\\[2mm]
\!\!\!\!\!\!\!&&\,\,\,- ab\left|\left\langle \tau_+\right.\left| \tau_-\right\rangle\right|\left( 
e^{-i\phi-i\alpha}\left| + \right\rangle \left\langle - \right|
+e^{i\phi+i\alpha}\left| - \right\rangle \left\langle + \right|\right) \,,\qquad\label{rho-1}
\end{eqnarray}
where we have used $\left\langle \tau_+\right.\left| \tau_-\right\rangle=\left|\left\langle \tau_+\right.\left| \tau_-\right\rangle\right|e^{i\alpha}.$

By resorting to the computations already made explicit in the previous Subsection, we arrive at the following expression for the detection probabilities:
\begin{eqnarray}
P_{\sigma_x=\pm1}=\frac{1}{2}\pm ab \left|\left\langle \tau_+\right.\left| \tau_-\right\rangle\right|\cos(\phi+\alpha)   \,,\label{P-composite}
\end{eqnarray}
which translates into the formula
\begin{eqnarray}
\mathcal{V}=2ab\left|\left\langle \tau_+\right.\left| \tau_-\right\rangle\right|\,.\label{visib-composit}
\end{eqnarray}
As for the probabilities $P_{\sigma_z=\pm1}$ and $\mathcal{P},$ we essentially recover the same result of Eq.~\eqref{predict-pure}.
Clearly, if $\left| \tau_\pm\right\rangle$ are orthogonal, then $\mathcal{V}=0$, which entails that, by measuring $\mathbb{1}_1\otimes 
\sigma_z,$ one could have a maximal access to the
which-way information. However, it is opportune to emphasize that the reduction of
visibility $\mathcal{V}$ does not depend on whether the measurements at $D_\pm$ have been carried out or not, and thus whether the two paths are \textit{distinguishable}. 
On the other hand, if the states $\left| \tau_\pm\right\rangle$  are not
orthogonal, then not even in principle the two paths can be perfectly distinguished.

We can better formalize the notion of distinguishable paths by introducing the \textit{distinguishability} as the \textit{trace norm distance}\footnote{The trace norm distance between two matrices (operators) $\sigma$ and  $\rho$ is defined as $T(\sigma,\rho):=\sqrt{(\sigma-\rho)^\dagger(\sigma-\rho)}.$ In the case of density matrices, $\sigma$ and $\rho$ are hermitian (but not necessarily positive) and hence $T(\sigma,\rho)=\frac{1}{2}\sqrt{(\sigma-\rho)^2}=\frac{1}{2}\sum_i \lambda_i,$ with $\lambda_i$ being the eigenvalues of the difference matrix $\sigma-\rho.$} between the final states of WWD, that is
\begin{eqnarray}
\mathcal{D}:=\frac{1}{2}{\rm Tr}\left\lbrace \left| \tau_+ \right\rangle  \left\langle \tau_+  \right| -\left| \tau_- \right\rangle  \left\langle \tau_-  \right|  \right\rbrace \,.\label{distinguish}
\end{eqnarray}
For the setup analyzed in Fig.~\ref{fig2}, it can be shown that\footnote{For a detailed derivation which includes the treatment of mixed states (for which the equality is replaced by an inequality $\leq 1$), see Ref.~\cite{Englert:1996zz}.}
\begin{eqnarray}
\mathcal{D}=\sqrt{1-\left|\left\langle \tau_+\right.\left| \tau_-\right\rangle\right|^2}\,.\label{distinguish-our}
\end{eqnarray}
In a nutshell, the distinguishability $\mathcal{D}$ of the two paths is the probability to correctly guess which path was taken by making a measurement on the subsystem WWD. 

At this point, by combining Eqs.~\eqref{visib-composit} and~\eqref{distinguish-our} and using the inequalities $\mathcal{D}, \mathcal{P}\leq1,$ we obtain 
\begin{eqnarray}
\mathcal{D}^2+\mathcal{V}^2=1 -(1-\mathcal{D}^2)\mathcal{P}^2\leq 1\,.\label{D+V<1}
\end{eqnarray}
We can now understand the physical meaning of the two inequalities~\eqref{P+V<1} and~\eqref{D+V<1}. 
\begin{itemize}

\item If $\mathcal{P}\neq 0,$ we can have access to the which-way information regardless of the outcome contained in the which-way detectors, namely even if $\mathcal{D} = 0$ the visibility $\mathcal{V}$ has to be limited;

\item if $\mathcal{P}=0,$  then $\mathcal{D}^2+\mathcal{V}^2=1$ for pure states (while for mixed states one has $\leq 1$), i.e. although the predictability is zero, the presence of WWD implies that the interferometric visibility $\mathcal{V}$ cannot be maximal
because $\mathcal{D}$ is non-vanishing.  

\end{itemize}
Let us conclude this Subsection by summarizing what we have learned about composite systems
with multiple degrees of freedom. The main message is that the complementarity principle holds also for composite systems.
Even if one manages to encode the information in the correlations between the subsystems, these quantum correlations necessarily correspond to entanglement, which means that the state of each subsystem is mixed and so
$\mathcal{V}$ cannot be maximal. If the subsystems were pure states, there would not be any entanglement between them and no
quantum correlation that could in principle reveal which-way information (in this case we would have had $\mathcal{D}=0$). 
Finally, let us remark again that the reduction or loss of visibility does not depend on whether the degrees of freedom encoded in WWD are measured or not.

Since we have reviewed the concept of quantum complementarity and the physical properties of the interferometric setup under consideration, we are ready to study the phenomenon of quantum interference in an external gravitational field. 

\section{Quantum interference of massive quantum clocks}\label{massive clock-sec}

In this Section, we will analyze a physical setup in which a Mach-Zehnder interferometer is embedded in a weak gravitational field, as for instance the one belonging to Earth (see Fig.~\ref{fig3}). As we will see below, in such a setting the presence of a non-vanishing gravitational potential induces a phase shift in the final state detected at $D_\pm.$ Furthermore, GR effects are responsible for a time dilation between the two arms of the interferometer, as they are located at two different heights with respect to Earth's surface. Quantum clock interferometry has been intensively studied in the context of Einstein's GR~\cite{Zych:2011hu,Pikovski:2013qwa,Sinha:2011mp,Dimopoulos:2008hx,Ufrecht:2020yay,Pumpo:2021pgv}.

Before continuing, let us point out that in this case the trajectories through the interferometer are supported against gravity. Although it is important to specify the mechanism with which the levitation is achieved (e.g. with a harmonic trap), here we just assume that a similar configuration exists and leave the aspects related to the precise experimental details of the apparatus for future works.

In this scenario, the r\^ole of WWD is played by time dilation effects associated with the \textit{internal} degrees of freedom of the interfering system. Therefore, we assume that $\left| \tau_{1,2}\right\rangle$ are internal states that work as clocks, where $1$ and $2$ refer to the upper and lower arm of the interferometer, respectively.  Thus, we consider a quantum version of the time dilation phenomenon in which a single clock is superposed along two paths having different proper times because of the gravitational potential of Earth. Remarkably, we are dealing with a quantum version of the \textit{twin paradox} with only a ``quantum child'' whose ``ages'' (proper times) are superposed, so that he/she is becoming older and younger than himself/herself at the same time.

Let us now  understand what happens in an interference experiment with a similar clock system.
The states $\left| +\right\rangle$, $\left| -\right\rangle$ related to the external degrees of freedom are now labeled by $\left| \gamma_{1}\right\rangle$,  $\left| \gamma_{2}\right\rangle$, where 
$\gamma_1$ and $\gamma_2$ denote the two paths traveled by the system in the interferometric apparatus (see Fig.~\ref{fig3}). We suppose that $\gamma_1$ lies farther from Earth's surface than $\gamma_2,$ so that the former feels a weaker gravitational potential with respect to the latter. 

Following the mathematical formalism introduced in the previous Section, we can write the quantum state of the \textit{composite clock system} inside the interferometer as follows~\cite{Zych:2011hu,Zych-thesis}
\begin{eqnarray}
\left| \psi \right\rangle=\frac{1}{\sqrt{2}}\left(e^{-i\phi_1}\left| \gamma_1 \right\rangle\left| \tau_1 \right\rangle +e^{-i\phi_2+i\phi}\left| \gamma_2 \right\rangle\left| \tau_2\right\rangle\right)\,,\label{clock-system-state}
\end{eqnarray}
where the phases $\phi_{1,2}$ are path-dependent and their values are intimately connected with the dynamics of the internal clock, whereas $\phi$ is some controllable phase shift. For the sake of simplicity, we work in the context of zero predictability, since $a=b=1/\sqrt{2}\Rightarrow \mathcal{P}=0.$ 

As already mentioned before, by reasonably accounting for the gravitational interaction on the clock in superposition, the internal degrees of freedom must evolve with different rates (i.e. proper times) along each path because of relativistic effects. This implies that such degrees of freedom enclose the information about which path is taken, thus acting as WWD.  

\begin{figure}[t]
	\includegraphics[scale=0.225]{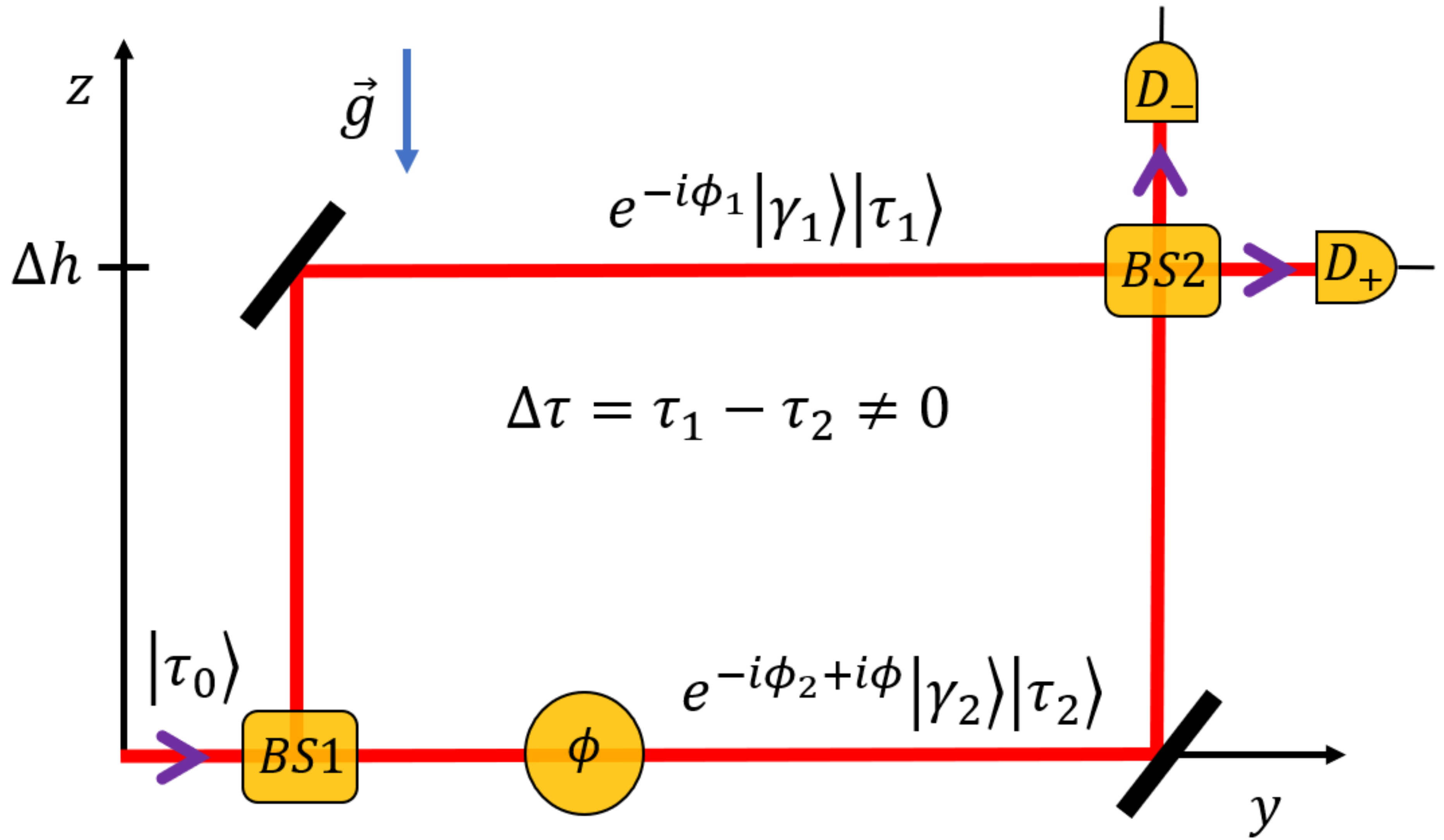}
	\centering
	\protect\caption{This figure illustrates the Mach-Zehnder interferometer in Fig.~\ref{fig1} placed in an external gravitational field, i.e. under the influence of a gravitational acceleration $\vec{g}$. The information about the which-way detector WWD is now encoded in an internal time-evolving degree of freedom that works as a clock. Due to time dilation effects, an initial internal state $\left| \tau_{0}\right\rangle$ can evolve into a linear combination of the two states $\left| \tau_{1,2}\right\rangle$ characterized by two different proper times, $\Delta\tau=\tau_1-\tau_2\neq 0.$ The phases $\phi_{1,2}$ depend on the path and on the dynamics of the internal degree of freedom, whereas $\phi$ is a controllable phase shift.}\label{fig3}
\end{figure}

As we have chosen the predictability to be zero, we can quantify the quantum complementarity by exploiting the concepts of interferometric visibility~\eqref{visib-composit} and distinguishability~\eqref{distinguish-our}, which in this case read
\begin{eqnarray}
\mathcal{V}=\left|\left\langle \tau_1\right.\left| \tau_2\right\rangle\right|\,,\label{visib-grav}
\end{eqnarray}
and 
\begin{eqnarray}
\mathcal{D}=\sqrt{1-\left|\left\langle \tau_1\right.\left| \tau_2\right\rangle\right|^2}\,,\label{disting-grav}
\end{eqnarray}
respectively. Furthermore, $P_{\sigma_z=\pm1}=1/2;$ instead, by following the steps contained in Sec.~\ref{composite-sec}, one can show that the probabilities to detect an outcome $\pm1$ at $D_\pm$ are given by
\begin{eqnarray}
P_{\sigma_x=\pm1}=\frac{1}{2}\pm \frac{1}{2}\left|\left\langle \tau_1\right.\left| \tau_2\right\rangle\right|\cos(\Delta\phi+\phi+\alpha)\,,\label{P-composit-grav}
\end{eqnarray}
where $\Delta\phi:=\phi_1-\phi_2$ and $\left\langle \tau_1\right.\left| \tau_2\right\rangle=\left|\left\langle \tau_1\right.\left| \tau_2\right\rangle\right|e^{i\alpha}.$

Let us recall one more time that the visibility is limited as a consequence
of the complementarity principle; indeed (for
total pure states) $\mathcal{V}^2 + \mathcal{D}^2 = 1$. We now note that, if the system is in a stationary internal state (namely, if the clock is switched off), then the which-way information is zero and $\mathcal{V}$ becomes
maximal. In fact, such a complementarity induced by time dilation can only be observed with a non-stationary internal state that is allowed to evolve along the two paths (i.e. with a switched-on clock).
Another comment is in order here: the distinguishability of the final states of the clock strictly depends on whether the difference $\Delta\tau:=\tau_1-\tau_2$ is larger or smaller than the time interval required by the internal states to evolve between two consecutive distinguishable states, that is the so-called \textit{orthogonalization time} $t_\perp$~\cite{Zych:2011hu,Zych-thesis}. 

In the next Subsections, we will derive the Hamiltonian of both the internal and the external degrees of freedom. Specifically, for the internal dynamics we will consider a two-level clock system as a model and perform precise computations of detection probabilities~\eqref{P-composit-grav}, interferometric visibility~\eqref{visib-grav} and distinguishability~\eqref{disting-grav} in the presence of a generic external, static and weak gravitational field.

\subsection{Hamiltonian of a quantum system in a gravitational field} \label{metric-hamilt-sec}

To comply with the above requests, in addition to the previous assumptions we also work under the condition that the evolution of the composite quantum system can be described in a low-energy regime, and that the relative distances among the internal constituents are
sufficiently small so as to neglect the variations of the metric over the size of the system.
This hypothesis conveys the idea that we can assign a single \textit{position} degree of freedom to the center-of-mass of the system. 
Consequently, the quantum system can be described as a point-like object with internal degrees of freedom, and we can still define a world line along which the proper time is measured. 

By accounting for these considerations, the total Hilbert space of the system is given by $\mathcal{H}=\mathcal{H}_{\rm ext}\otimes\mathcal{H}_{\rm int},$ where $\mathcal{H}_{\rm ext}$ includes states related to external (center-of-mass) degrees of freedom, whilst $\mathcal{H}_{\rm int}$ incorporates the states describing internal degrees of freedom, or in other words the states of the clock.
Concerning the gravitational sector, the spacetime background can be described by a generic linearized static metric expressed in isotropic coordinates
\begin{eqnarray}
\!{\rm d}s^2\!=\!-\left(1+\frac{2\Phi}{c^2}\right)c^2{\rm d}t^2\!+\!\left(1-\frac{2\Psi}{c^2}\right)\!({\rm d}r^2\!+\!r^2{\rm d}\Omega^2),\,\,\label{metric}
\end{eqnarray}
where $r=\sqrt{x^2+y^2+z^2},$ $c$ is the speed of light  and $\Phi(r)$ and $\Psi(r)$ are two generic metric potentials; in the case of GR, we have $\Phi=\Psi=-GM/r,$ with $G$ being the Newton's constant and $M$ the mass of the source (in our case $M=M_\oplus$).

Moreover, for simplicity and for consistency with Refs.~\cite{Zych:2011hu,Zych-thesis}, as a preliminary investigation we neglect any complexity stemming from the spinor nature of the examined system; therefore, we study a real scalar field $\varphi$ of rest mass $m_r$ in curved spacetime, whose field equation is given by 
\begin{eqnarray}
\left(\Box_g-\frac{m_r^2c^2}{\hbar^2}\right)\varphi(x)=0\,,\label{klein-g-curved}
\end{eqnarray}
where  $\Box_g=g^{\mu\nu}\nabla_\mu\nabla_\nu$ is the curved d'Alembertian, which acts on scalar quantities as follows:
\begin{eqnarray}
\Box_g \varphi(x)=\frac{1}{\sqrt{-g}}\partial_\mu\left(\sqrt{-g}g^{\mu\nu}\partial_\nu\right)\varphi(x)\,.\label{d'alembert-scalars}
\end{eqnarray}
In the linearized regime (i.e. up to linear order in $G$), by using the form of the metric given in Eq.~\eqref{metric} we obtain
\begin{eqnarray}
&&\Big(\Box+2c^{-2}\Phi\partial_0^2+2c^{-2}\Psi\nabla^2\nonumber\\
&&\qquad+c^{-2}\vec{\nabla}(\Phi-\Psi)\cdot\vec{\nabla}-\frac{m_r^2c^2}{\hbar^2}\Big)\varphi(x)=0\,,\,\,\label{klein-g-curved-lin}
\end{eqnarray}
where $\Box=-\partial_0^2+\nabla^2$ is the flat d'Alembertian, with $\partial_0^2=c^{-2}\partial^2_t$ and $\nabla^2=\delta^{ij}\partial_i\partial_j.$

By thoroughly relying on the procedure introduced in Refs.~\cite{Kiefer:1990pt,Lammerzahl:1995zz}, we perform a non-relativistic expansion to compute the Schr\"odinger equation of a quantum system in the external spacetime metric~\eqref{metric}
\begin{eqnarray}
i\hbar\partial_t \psi(t,\vec{x})=H\psi(t,\vec{x})\,,\label{schr}
\end{eqnarray}
where $\psi(t,\vec{x})$ represents the quantum wave function and $H$ is the Hamiltonian 
\begin{eqnarray}
\!H\!\!&=& \!\!m_rc^2+\frac{p^2}{2m_r}-\frac{p^4}{8m_r^3 c^2}+m_r\Phi+\frac{1}{m_rc^2}\left(\frac{\Phi}{2}+\Psi\right)p^2\nonumber\\[2mm]
&& \!-\frac{1}{4m_rc^2}[p^2\Phi]-\frac{1}{2m_rc^2}[\vec{p}\,\Psi]\cdot\vec{p} \,,\,\,\,\quad\label{Hamiltonian}
\end{eqnarray}
where $[\cdots]$ indicates that the momentum operator acts only on the object inside the brackets. In the case of GR, as $\Psi$ and $\Phi$ are equal to the Newtonian potential, we recover the result contemplated in Ref.~\cite{Lammerzahl:1995zz}. The algebraic details on the derivation of Eqs.~\eqref{schr} and~\eqref{Hamiltonian} can be found in Appendix~\ref{schr-curved-sec}.

Let us emphasize that the Hamiltonian~\eqref{Hamiltonian} acts on the product space $\mathcal{H},$ thereby governing the dynamics of both the external and the internal degrees of freedom of the composite quantum system. 
Another crucial aspect to pinpoint is that in general the rest mass $m_r$ contains \textit{two} distinct contributions~\cite{Zych:2011hu}, which are
\begin{eqnarray}
m_r=m\mathbb{1}_{\rm int}+c^{-2}H_0\,,\label{Hamilt-internal}
\end{eqnarray}
where $m$ is the static rest mass, $c^{-2}H_0$ is the dynamical contribution
due to the internal degrees of freedom and $\mathbb{1}_{\rm int}$ is the identity operator on  $\mathcal{H}_{\rm int}.$

If we ignore higher-order terms in $c^{-2},$ we can write the Hamiltonian~\eqref{Hamiltonian} in the following compact form:
\begin{eqnarray}
H=H_{\rm cm}+H_0\left(1+\frac{\Gamma(x,p)}{c^2}\right)\,,\label{Hamilt-compact}
\end{eqnarray}
with 
\begin{eqnarray}
H_{\rm cm}:= mc^2+\frac{p^2}{2m}+m\Phi+H_{\rm corr}\,,\quad\label{Hamiltonian-cm}
\end{eqnarray}
and
\begin{eqnarray}
\Gamma(x,p):=\Phi(x)-\frac{p^2}{2m^2}\,.\quad\label{gamma}
\end{eqnarray}
The term $H_{\rm corr}$ includes special and general relativistic corrections that will correspond to a mere phase shift in the interference patter, which is not relevant for our purposes and so we can exclude them henceforth. Instead, the term  $\Gamma(x,p)$ is extremely important and grants access to the time dilation effect.
It is worth stressing that, up to the considered approximation, the gravitational potential $\Psi$ does not contribute to the time dilation, but only to $H_{\rm corr};$ hence, in this regime its only implication results in an additional phase shift, which we do not show explicitly.

Now, using the expression for the Hamiltonian~\eqref{Hamilt-compact}, we can evaluate the evolution of the state inside the interferometer. First of all, let us look at the following total pure state:
\begin{eqnarray}
\left| \psi \right\rangle=\frac{1}{\sqrt{2}}\left(\left| \psi_1 \right\rangle +e^{i\phi}\left| \psi_2 \right\rangle\right)\,.\label{total-state}
\end{eqnarray}
The states $\left| \psi_{1,2} \right\rangle$ are associated with the two paths $\gamma_{1,2}$ and can be determined by acting with the evolution operator on the initial state $\left| x_{\rm in} \right\rangle\left| \tau_0 \right\rangle,$ thus yielding
\begin{eqnarray}
\left| \psi_i \right\rangle=e^{-\frac{i}{\hbar}\int_{\gamma_i}{\rm d}t\left[H_{\rm cm}+H_0\left(1+\frac{\Gamma}{c^2}\right)\right]}\left| x_{\rm in} \right\rangle\left| \tau_0 \right\rangle\,,\,\,\,\, i=1,2\,.\,\,\,\,\,\label{evolution-state}
\end{eqnarray}
Note that $\int_{\gamma_i}{\rm d}t(1+\Gamma/c^2)=\int_{\gamma_i} {\rm d}\tau_i,$ where $\tau_i$ is the proper time along the path $\gamma_i$. By further assuming that $H_0$ is time-independent, we can write
\begin{eqnarray}
\left| \psi_i \right\rangle&=&\left[e^{-\frac{i}{\hbar}\int_{\gamma_i}{\rm d}t H_{\rm cm}}\left| x_{\rm in} \right\rangle\right] \left[e^{-\frac{i}{\hbar}H_0\tau_i}\left| \tau_0 \right\rangle\right]\nonumber\\[2mm]
&\equiv& \left| \gamma_i\right\rangle\left| \tau_i \right\rangle\,,
\end{eqnarray}
where we have used
\begin{eqnarray}
\left| \gamma_i\right\rangle=e^{-\frac{i}{\hbar}\int_{\gamma_i}{\rm d}t H_{\rm cm}}\left| x_{\rm in} \right\rangle\,,
\end{eqnarray}
and 
\begin{eqnarray}
\left| \tau_i\right\rangle=e^{-\frac{i}{\hbar}H_0\tau_i}\left| \tau_0 \right\rangle\,.
\end{eqnarray}
From the above equation, it is straightforward to calculate the interferometric visibility~\eqref{visib-grav}, that is
\begin{eqnarray}
\mathcal{V}=|\left\langle \tau_1 \right.\left| \tau_2\right\rangle|= |\left\langle \tau_0 \right| e^{\frac{i}{\hbar}H_0\Delta\tau}\left| \tau_0\right\rangle|\,,\label{visibility-grav-H0}
\end{eqnarray}
where we have denoted $\Delta\tau=\tau_1-\tau_2$ as the proper time difference between the two paths $\gamma_1$ and $\gamma_2.$
The internal contribution to the Hamiltonian~\eqref{Hamilt-compact} not only is responsible for time dilation effects, but it also gives rise to entanglement between internal and external degrees of freedom. In turn, the interferometric visibility is not equal to one but it decreases, thereby signaling a loss of coherence.

By virtue of the scheme depicted so far, we have extended the results of Refs.~\cite{Zych:2011hu,Zych-thesis} to the case in which the spacetime metric is \textit{not} described by GR, as in general we may have $\Phi\neq \Psi\neq -GM/r.$ However, we have seen that $\Psi$ only contributes to an irrelevant phase shift, whereas $\Phi$ is the source of the dominant contribution to the phase shift and of time dilation. For each extended theory of gravity beyond Einstein's GR, there is a corresponding modified Newtonian potential $\Phi$ which will cause different phase shifts and losses of coherence.
In the next Subsection, we consider a two-level quantum system as a clock to find a simple expression for $H_0,$ by means of which we can explicitly evaluate the interferometric visibility~\eqref{visibility-grav-H0}.

\subsection{Two-level system as a quantum clock} \label{two-level-sec}

A two-level system is the most immediate setup we can think of to construct a quantum clock and to better understand how time dilation can induce a loss of coherence.
Let us assume that $\mathcal{H}_{\rm int}$ is two-dimensional and is spanned by the basis of two energy eigenstates $\left\lbrace \left| 1\right\rangle, \left| 2\right\rangle   \right\rbrace$ of the operator $H_0,$ with eigenvalues $E_1$ and $E_2,$ respectively. Therefore, we can cast the operator $H_0$ as
\begin{eqnarray}
H_0=E_1\left|1 \right\rangle \left\langle 1\right|+ E_2\left|2 \right\rangle \left\langle 2\right|\,.\label{H0}
\end{eqnarray}
In light of this, the initial internal state can be expressed as
\begin{eqnarray}
\left|\tau_0 \right\rangle=\frac{1}{\sqrt{2}}\left(\left|1 \right\rangle+\left|2 \right\rangle \right) \,,\label{in-int-state}
\end{eqnarray}
so that the evolved internal state is given by
\begin{eqnarray}
\left|\tau_i \right\rangle=\frac{1}{\sqrt{2}}\left(e^{-\frac{i}{\hbar}E_1\tau_1}\left|1 \right\rangle+e^{-\frac{i}{\hbar}E_2\tau_2}\left|2 \right\rangle \right) \,.\label{tau_i-int-state}
\end{eqnarray}
With this knowledge, we now have all the ingredients to explicitly evaluate the interferometric visibility for an initial pure state in Eq.~\eqref{visibility-grav-H0}, which gives
\begin{eqnarray}
\mathcal{V}= \left| \cos \left(\frac{\Delta E\Delta\tau}{2\hbar} \right)\right|\,,\label{visibility-grav-explicit}
\end{eqnarray}
where $\Delta E:=|E_2-E_1|.$ By introducing the orthogonalization time $t_\perp=\frac{\pi\hbar}{\Delta E}$~\cite{Zych:2011hu} we can rephrase the last equation in a different shape, that is
\begin{eqnarray}
\mathcal{V}= \left| \cos \left(\frac{\Delta\tau}{t_\perp}\frac{\pi}{2} \right)\right|\,.\label{visibility-grav-explicit-orthog-time}
\end{eqnarray}
As long as $\Delta\tau\geq t_\perp,$ there will be an amount of accessible which-way information encoded in the internal states of the interfering system, which is traduced in a loss of coherence.

In order to compute $\Delta\tau$, we recall that we have set the interferometer so that the gravitational acceleration is only present along the $z$-direction and that the velocity of the quantum system has only $y$-component. At this stage, we further make the reasonable ansatz according to which the velocities are the same for both paths, in such a way that the existence of time dilation is solely due to the presence of the external gravitational field. Now, by expanding the gravitational potential in the limit $\Delta h\ll R,$ with $\Delta h$ being the distance between the two arms of the interferometer and $R$ the Earth's radius as well as the height of the path $\gamma_2$ (see Fig.~\ref{fig3}), namely
\begin{eqnarray}
\Phi(R+\Delta h)\simeq \Phi(R)+\Phi^\prime(R)\Delta h\,,\label{pot-exp}
\end{eqnarray}
we can write
\begin{eqnarray}
\Delta \tau&=&c^{-2}\int_0^{\Delta T}{\rm d}\tau\left[\Phi(R+\Delta h)-\Phi(R)\right]\nonumber\\
&=&c^{-2}\Phi^\prime(R) \Delta h  \Delta T\,,
\end{eqnarray}
where $\Delta T$ is the coordinate \textit{travel time} as seen from the laboratory frame, which expresses the time exerted by the quantum system to travel through the two arms of the interferometer at two constant heights $R$ and $R+\Delta h.$ Note that we have used the shorthand notation $f^\prime\equiv {\rm d}f/{\rm d}r.$
Therefore, the interferometric visibility can be reformulated as
\begin{eqnarray}
\mathcal{V}= \left| \cos \left(\frac{\Phi^\prime (R) \Delta h\Delta T\Delta E}{2\hbar c^2} \right)\right|\,.\label{visibility-grav-deriv-pot}
\end{eqnarray}
Furthermore, we can also compute the probabilities $P_{\sigma_x=\pm1},$ which turn out to be equal to
\begin{eqnarray}
\!\!\!\!P_{\sigma_x=\pm1}&=&\frac{1}{2}\pm \frac{1}{2}\left| \cos \left(\frac{\Phi^\prime (R) \Delta h\Delta T\Delta E}{2\hbar c^2} \right)\right|\nonumber \\[2mm]
&&\times \cos\left(\frac{m\Phi^\prime(R)\Delta h \Delta T}{\hbar}+\phi+ \beta\right)\,,\label{Probab-grav-expl}
\end{eqnarray}
with $\beta$ being a phase which includes both special and general relativistic contributions and whose explicit form is not relevant for our purposes (see Ref.~\cite{Zych:2011hu} for its derivation in the context of GR).

To grasp the meaning of the behavior of the detection probabilities and of the
interferometric visibility, we can study the following difference
\begin{eqnarray}
&&P_{\sigma_x=+1}-P_{\sigma_x=-1}=\left| \cos \left(\frac{\Phi^\prime (R) \Delta h\Delta T\Delta E}{2\hbar c^2} \right)\right|\nonumber \\[2mm]
&&\qquad\quad \times \cos\left(\frac{m\Phi^\prime(R)\Delta h \Delta T}{\hbar}+\phi+ \beta\right)\,.\label{Probab-difference}
\end{eqnarray}
Since we are interested in theories beyond Einstein's GR, we can write the gravitational potential as
\begin{eqnarray}
\Phi(r)=\Phi_{\rm GR}(r)+\chi(r)\,,\label{corrected potential}
\end{eqnarray}
where $\Phi_{GR}=-GM/r$ while $\chi$ takes into account corrections to Newton's potential; the same separation can be done also for $\beta=\beta_{\rm GR}+\delta.$ 
The corrections due to new physics beyond GR affect both the interferometric visibility and the phase shift. This feature opens a new window of opportunities to test, constrain and discriminate extended theories of gravity in tabletop laboratory experiments based upon quantum interference. 


\subsection{Non-relativistic regime: COW effect}\label{cow sec}

\begin{figure}[t]
	\includegraphics[scale=0.505]{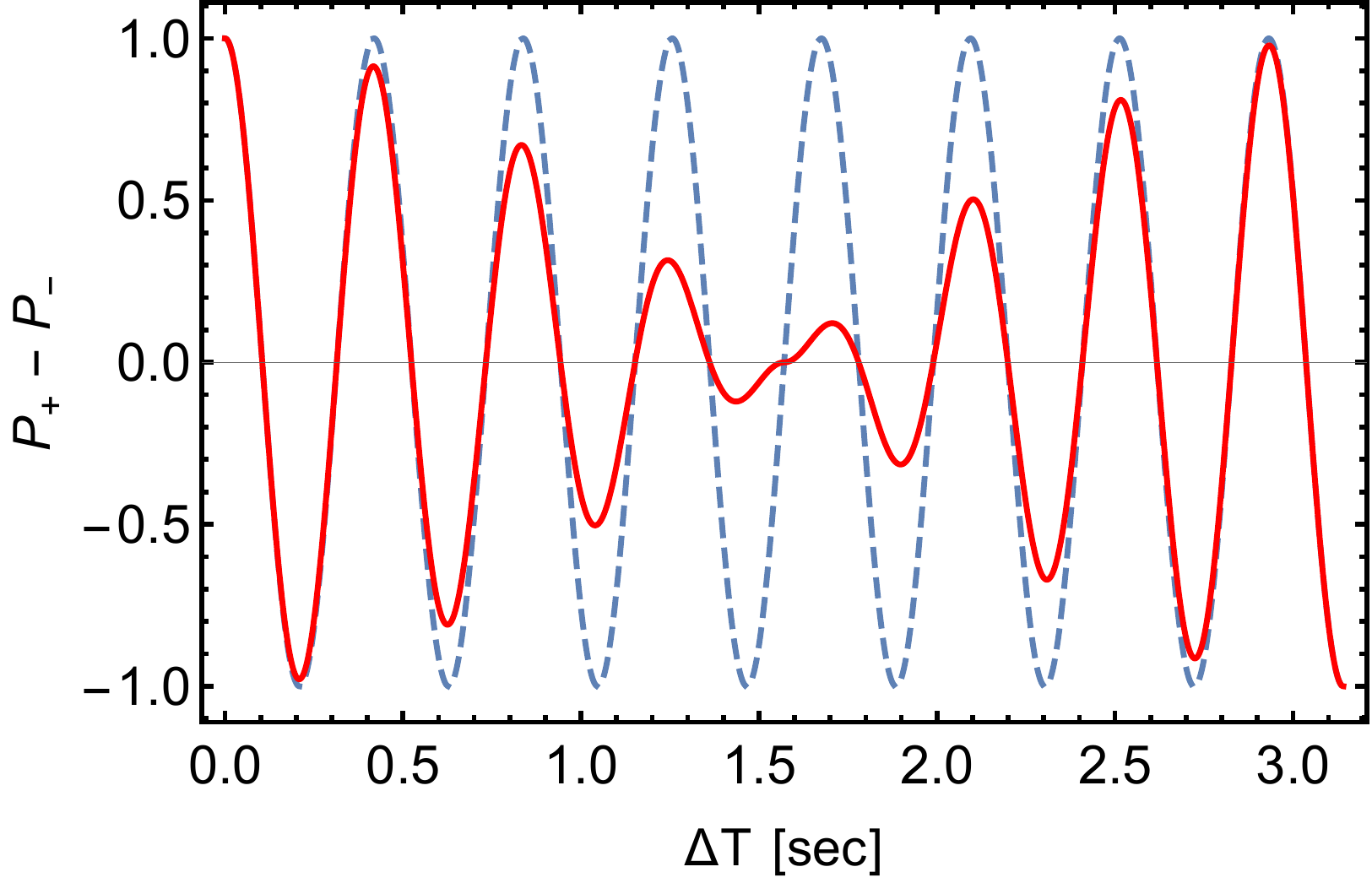}
	\centering
	\protect\caption{Quantum interference in GR. The blue dashed line corresponds to the coherent oscillation due to the COW effect, whereas the red solid line represents the interference pattern in the presence of time dilation. One can explicitly notice that, because of time dilation, there is a loss of coherence. For the sake of argument, we have set $\Phi^\prime\Delta h\Delta E/(2\hbar c^2)=1$ Hz,  $m\Phi^\prime\Delta h/\hbar=15$ Hz, $\phi=\beta=0.$}\label{fig4}
\end{figure}

In the non-relativistic regime, the term ${\Phi^\prime (R) \Delta h\Delta T\Delta E}/{2\hbar c^2} \sim \mathcal{O}(c^{-2})$ is negligible (i.e. no time dilation effect appears), thereby allowing the detection probability to take the form
\begin{eqnarray}
P^{\rm nr}_{\sigma_x=+1}-P^{\rm nr}_{\sigma_x=-1}= \cos\left(\frac{m\Phi^\prime(R)\Delta h \Delta T}{\hbar}+\phi\right)\,.\label{Probab-difference-non-relav}
\end{eqnarray}
The term $\Delta\varphi:={m\Phi^\prime(R)\Delta h \Delta T}/{\hbar}\sim \mathcal{O}(c^{0})$ is the non-relativistic phase shift due to the gravitational potential, and it has been measured for the first time in the COW experiment~\cite{Colella:1975dq}. The phenomenon is also known under the name ``COW effect'', and it can be regarded as the gravitational analogue of the
Bohm-Aharonov effect~\cite{Aharonov:1959fk}. This was the first-ever gravitational effect measured on a quantum system. The blue dashed curve in Fig.~\ref{fig4} portraits the qualitative behavior of the detection probabilities in the case of the COW effect.

The first measurement of COW effect~\cite{Colella:1975dq} was performed with neutrons, but many experimental improvements have been made in the last decades~\cite{Abele:2012dn}. To test and constrain gravitational theories beyond GR, we must compare the magnitude of the phase shift induced by the corrected Newtonian potential~\eqref{corrected potential} with the experimental error.
From Refs.~\cite{Staudenmann:1980uqe,thesis-cow}, one can see that the current experimental error $\Delta_{\rm err} \varphi$ on the determination of the gravitational phase shift is of the order of
\begin{eqnarray}
\Delta_{\rm err} \varphi=\pm0.001\,\,{\rm rad}\,.\label{exp-error-cow}
\end{eqnarray}
Thus, for consistency with the experimental data, any predicted correction $\Delta_{\chi}\varphi$ to the phase shift induced by alternative theories beyond Einstein's GR such that $\Delta\varphi=\Delta_{\rm GR}\varphi+\Delta_{\chi}\varphi$ must satisfy the following constraint:
\begin{eqnarray}
|\Delta_{\chi} \varphi|<|\Delta_{\rm err} \varphi|=0.001\,\,{\rm rad}\,.\label{constraint-inequality}
\end{eqnarray}
%

\subsection{General relativistic regime: time dilation}\label{time-dil sec}

In the fully relativistic regime, the term ${\Phi^\prime (R) \Delta h\Delta T\Delta E}/{2\hbar c^2} \sim \mathcal{O}(c^{-2})$ is no longer negligible, thereby giving rise to drastic differences with respect to the non-relativistic scenario. Because of this term, the absolute value of the cosine function~\eqref{Probab-difference} is fundamental, as it physically causes a loss of coherence as depicted in Fig.~\ref{fig4}.

This intriguing effect has not been observed yet in a purely gravitational experiment. However, in Ref.~\cite{Margalit} the very same decoherence induced by time dilation for a two-level system was experimentally simulated with a Bose-Einstein condensate. In this test, the authors consider a Stern-Gerlach type matter-wave interferometer under the influence of an external inhomogeneous magnetic field which mimics the effect of gravitational time dilation. In a similar framework, the quantum clock is simply represented by the spin precession of the system. The final statement of Ref.~\cite{Margalit} asserts that the claimed result may potentially lead to a deeper study of self-interacting clocks in a laboratory on Earth.
Moreover, there are good chances that, in the near future, it will be possible to achieve the high sensitivity required to perform a proper gravitational experiment~\cite{Zych:2011hu}.


\section{Extended theories of gravity beyond General Relativity}\label{quant-int-beyond-sec}

In this Section, we introduce a wide class of alternative theories of gravity whose action contains quadratic curvature terms in addition to the Einstein-Hilbert part. We show the expressions for the corresponding gravitational potentials in the weak-field approximation, which will be employed to evaluate the relevant physical quantities for our quantum interference setup.
In particular, we start from the following generic gravitational action \cite{Biswas:2005qr,Biswas:2016etb}:
\begin{eqnarray}
\nonumber
S&\hspace{-1mm}=\hspace{-1mm}& \frac{1}{2\kappa^2}\int d^4x\sqrt{-g}\left\lbrace \mathcal{R}+\frac{1}{2}\Big[\mathcal{R}F_1(\Box_g)\mathcal{R}\right.\\[2mm]
&&\qquad\qquad+\,\mathcal{R}_{\mu\nu}F_2(\Box_g)\mathcal{R}^{\mu\nu}\Big]\bigg\rbrace,
\label{quad-action}
\end{eqnarray}
where $\kappa\equiv \sqrt{8\pi G/c^4},$ and the differential operators $F_i(\Box_g)$ can be either analytic or non-analytic functions of $\Box_g$
\begin{equation}
F_i(\Box_g)=\sum\limits_{n=0}^{N}f_{i,n}\Box_g^n,\,\,\,\,\,\,\,\,\,i=1,2.
\end{equation}
Positive (negative) powers of the d'Alembertian, that is, $n>0$ ($n<0$), correspond to ultraviolet (infrared) generalizations of Einstein's GR. When $N<\infty$  and $n>0$, we have a local (polynomial) theory of gravity, whose derivative order is $2N+4$; when $N=\infty$ and/or $n<0$ we have a nonlocal (non-polynomial) theory of gravity, which means that the two form factors $F_i(\Box_g)$ can be non-polynomial differential operators of $\Box_g.$

In what follows, we  list several theories of gravity and the ensuing modified Newtonian potentials. Let us recall that we deal with static and spherically symmetric backgrounds; for this reason, the general form of the linearized spacetime metric we investigate is the one given by Eq.~\eqref{metric}.
In Appendix~\ref{gen-poisson-sec}, we review quadratic theories of gravity and their linearized regime more accurately, including a derivation of the modified Poisson equations for the metric potentials.

\subsection{$\mathcal{R}^2$-gravity}

Let us first start from the simplest quadratic extension of the Einstein-Hilbert action which involves a Ricci scalar squared term with constant form factor
\begin{equation}
F_1=\alpha\,,\qquad F_2=0\,.
\end{equation}
This choice belongs to the wide class of $f(\mathcal{R})$-theories, where the Lagrangian can be a generic function of the Ricci scalar. 

For such a model, the two metric potentials $\Phi$ and $\Psi$ defined in Eq.~\eqref{metric} are 
\begin{eqnarray}\label{fr}
\nonumber
\Phi(r)&=&\displaystyle -\frac{GM}{r}\left(1+\frac{1}{3}e^{-m_0r}\right),\\[2mm]
\Psi(r)&=&\displaystyle -\frac{GM}{r}\left(1-\frac{1}{3}e^{-m_0r}\right),
\end{eqnarray}
with $m_0=1/\sqrt{3\,\alpha}$ being the inverse of a length and the mass of an extra massive spin-$0$ degree of freedom.

The potential $\Phi$ is the only term that appears in the regime we are interested in (see Eq.~\eqref{Probab-difference}). From its expression, we can compute the correction $\chi(r)$ to the Newtonian potential introduced in Eq.~\eqref{corrected potential}, which reads
\begin{eqnarray}
\chi(r)=-\frac{GM}{3r}e^{-m_0r}\,.
\end{eqnarray}

\subsection{Stelle gravity}

For Stelle four-derivative gravity~\cite{-K.-S.}, the differential operators of the action are chosen to be
\begin{equation}
F_1=\alpha\,,\qquad F_2=\beta\,.
\end{equation}
Consequently, the two metric potentials are
\begin{eqnarray}
\nonumber
\Phi(r)&=& -\frac{GM}{r}\left(1+\frac{1}{3}e^{-m_0r}-\frac{4}{3}e^{-m_2r}\right),\\[2mm]
\Psi(r)&=& -\frac{GM}{r}\left(1-\frac{1}{3}e^{-m_0r}-\frac{2}{3}e^{-m_2r}\right),\label{stelle-pot}
\end{eqnarray}
where $m_0=2/\sqrt{12\,\alpha+\beta}$ and $m_2=\sqrt{2/(-\beta)}$ are the inverse of a length and the masses of an extra spin-$0$ and a spin-$2$ massive degree of freedom, respectively. Note that we must impose $\beta<0$ in order to avoid tachyonic modes.

In this case, the correction $\chi(r)$ is
\begin{eqnarray}
\chi(r)=-\frac{GM}{3r}e^{-m_0r}+\frac{4GM}{3r}e^{-m_2r}\,.
\end{eqnarray}

\subsection{Analytic nonlocal gravity}

As an analytic nonlocal theory of gravity, we select the following form factors~\cite{Biswas:2011ar}:
\begin{equation}
F_1=-\frac{1}{2}F_2=\,\frac{1-e^{-\ell^2\Box_g}}{2\,\Box_g}\,, \label{ghost-free-choice}
\end{equation}
where $\ell$ is the fundamental length scale of nonlocality at which new gravitational physics should become manifest. For this theory, the only propagating degree of freedom around Minkowski background is the massless transverse spin-$2$ graviton with $\pm2$ helicities.

Due to the peculiar choice for $F_1$ and $F_2$, the two metric potentials coincide and are given by
\begin{equation}
\Phi(r)=\Psi(r)=-\frac{GM}{r}\,{\rm Erf}\left(\frac{r}{2\ell}\right),\label{ghost-free-pot}
\end{equation}
where ${\rm Erf}(x)=\frac{2}{\sqrt{\pi}}\int^x_0e^{-t^2}dt$ is the so-called error function.

Now, the corresponding correction to the Newtonian gravitational field is
\begin{eqnarray}
\chi(r)=\frac{GM}{r}{\rm Erfc}\left(\frac{r}{2\ell}\right)\,,
\end{eqnarray}
where ${\rm Erfc}(x)=1-{\rm Erf}(x)$ is the complementary error function.

\subsection{Non-analytic nonlocal gravity}

Finally, we consider a non-analytic nonlocal model which is realized with the following choice of form factors:
\begin{equation}
F_1=\frac{\alpha}{\Box}\,,\qquad F_2=0\,.\label{nonlocal-choice1}
\end{equation}
The two metric potentials contain an infrared modification of the standard Newtonian one, as it can be seen from their expression, namely
\begin{eqnarray}\label{pono}
\nonumber
\Phi(r)&=& -\frac{GM}{r}\left(\frac{4\alpha-1}{3\alpha-1}\right),\\[2mm]
\Psi(r)&=&\displaystyle -\frac{GM}{r}\left(\frac{2\alpha-1}{3\alpha-1}\right).
\end{eqnarray}
In this case, the correction to the metric potential reads
\begin{eqnarray}
\chi(r)=-\frac{\alpha}{3\alpha - 1}\frac{GM}{r}\,.
\end{eqnarray}

\section{Discussion \& experimental constraints}\label{discuss-sec}

After the calculation of the quantity $\chi(r)$ for different quadratic models of gravity, we can now compute the detection probability~\eqref{Probab-difference} for each gravitational theory to understand how quantum interference is affected by the presence of a corrected Newtonian potential with respect to standard GR. To this aim, we work with both the non-relativistic COW effect and the relativistic decoherence arising from time dilation to constrain and explore (classical) modified gravity via its interplay with quantum mechanics.

Let us start from the COW effect and recall that the corrections $\chi(r)$ to the Newtonian potential must satisfy the experimental bound~\eqref{constraint-inequality}, which can be rephrased as
\begin{eqnarray}
|\chi^{\prime}(R_\oplus)|<(0.001)\times \frac{\hbar}{m\Delta h\Delta T}\,,\label{ineq-constr}
\end{eqnarray}
where $R_\oplus$ is Earth's radius, $m$ is the mass of the interfering quantum system, $\Delta h$ is the distance between the two arms of the interferometer and $\Delta T$ is the travel time, or in other words the duration of the quantum superposition.

In what follows, we consider neutrons as the superposed quantum system for which the spin precession acts as a clock; see Ref.~\cite{Zych:2011hu} for more details on the experimental apparatus and data. By using the neutron mass $m\simeq 1.67\times 10^{-27}$kg, $\hbar\simeq 1.05\times 10^{-34}\,{\rm m^2\,kg/s}$ and the achieved experimental value $\Delta h\Delta T\simeq 10^{-6}\,{\rm m\cdot s},$ we can rewrite the bound~\eqref{ineq-constr} as
\begin{eqnarray}
|\chi^{\prime}(R_\oplus)|<6.5\times 10^{-5}\,{\rm \frac{m}{s^2}}\,.\label{ineq-constr-2}
\end{eqnarray}
By virtue of this bound, we can constrain the gravitational theories introduced in the previous Section using $R_\oplus\simeq 6.37\times 10^6$ m and $GM_\oplus\simeq 4\times 10^{14}$ ${\rm m^3/s^2}.$
\begin{itemize}
	
	\item In the case of $\mathcal{R}^2$-gravity, the inequality reads
	%
	\begin{eqnarray}
	e^{-R_\oplus m_0}(1+R_\oplus m_0)\lesssim 6.4\times 10^{-6}\,,\label{ineq-constr-R^2}
	\end{eqnarray}
	from which we get the following constraint on the free parameter
	\begin{eqnarray}
		\frac{1}{m_0}\lesssim 4.3\times 10^{5}\,.\label{ineq-constr-R^2-lambda}
	\end{eqnarray}

	\item For Stelle gravity, we have
	\begin{eqnarray}
	&&\left|e^{-R_\oplus m_0}(1+R_\oplus m_0)\right.\nonumber\\
	&&\qquad\quad\left.-4e^{-R_\oplus m_2}(1+R_\oplus m_2)\right|\lesssim 6.4\times 10^{-6}\,,\label{ineq-constr-stelle}
	\end{eqnarray}
	from which we obtain constraints on $m_0$ and $m_2$ of the same order of the one in Eq.~\eqref{ineq-constr-R^2-lambda}.

    \item In the context of analytic nonlocal gravity, the constraint yields
    \begin{eqnarray}
    \left|\frac{e^{-R_\oplus^2/4\ell^2}R_\oplus}{\sqrt{\pi}\ell}+{\rm Erfc}\left(\frac{R_\oplus}{2\ell}\right)\right|\lesssim 6.4\times 10^{-6}\,,\label{ineq-constr-IDG}
    \end{eqnarray}
    from which it follows
    \begin{eqnarray}
    	\ell\lesssim 8.7\times 10^{5}\,.\label{ineq-constr-IDG-lambda}
    \end{eqnarray}

    \item Finally, in the framework of non-analytic nonlocal gravity we obtain
    \begin{eqnarray}
    \left|\frac{\alpha}{3\alpha-1}\right|\lesssim 6.4\times 10^{-6}\,,\label{ineq-constr-non-analytic}
    \end{eqnarray}
    and the constraint on the free parameter reads
    \begin{eqnarray}
    	|\alpha|\lesssim 6.4\times 10^{-6}\,.\label{ineq-constr-alpha-lambda}
    \end{eqnarray}

\end{itemize}

\noindent
Interestingly, these constraints are of the same order of the ones coming from Gravity Probe B~\cite{Everitt:2015qri}, which is the best satellite experiment so far. On the other hand, we should point out that the best laboratory constraint on deviations from Newton’s law still comes from torsion-balance experiments performed on Earth. To give an idea on the precise order of magnitude, we can rely on the results coming from the E\"ot-Wash experiment, which gives $1/m_0,\,\ell\lesssim 10\,{\rm \mu m}$~\cite{Kapner:2006si}.


It is common belief that, in the near future, a huge development will be made in this sector of quantum interference by bringing heavier systems in superposition and by simultaneously increasing the travel time (i.e. the length of the arms). For instance, there are promising indications towards the feasibility of superposing heavy masses of the order of $10^{-16}$kg (see Ref.~\cite{Bose:2017nin} for related discussions) and at the same time achieving values $\Delta h\Delta T\simeq 10^{6}$ ${\rm m\cdot s}$ \cite{Zych:2011hu}. Together with a smaller experimental error $\Delta_{\rm err}\varphi,$ this technological enhancement would allow us to significantly decrease the magnitude of the r.h.s. of the inequality constraints~(\ref{ineq-constr-R^2-lambda},\ref{ineq-constr-IDG-lambda},\ref{ineq-constr-alpha-lambda}).

\begin{figure}[t]
	\includegraphics[scale=0.385]{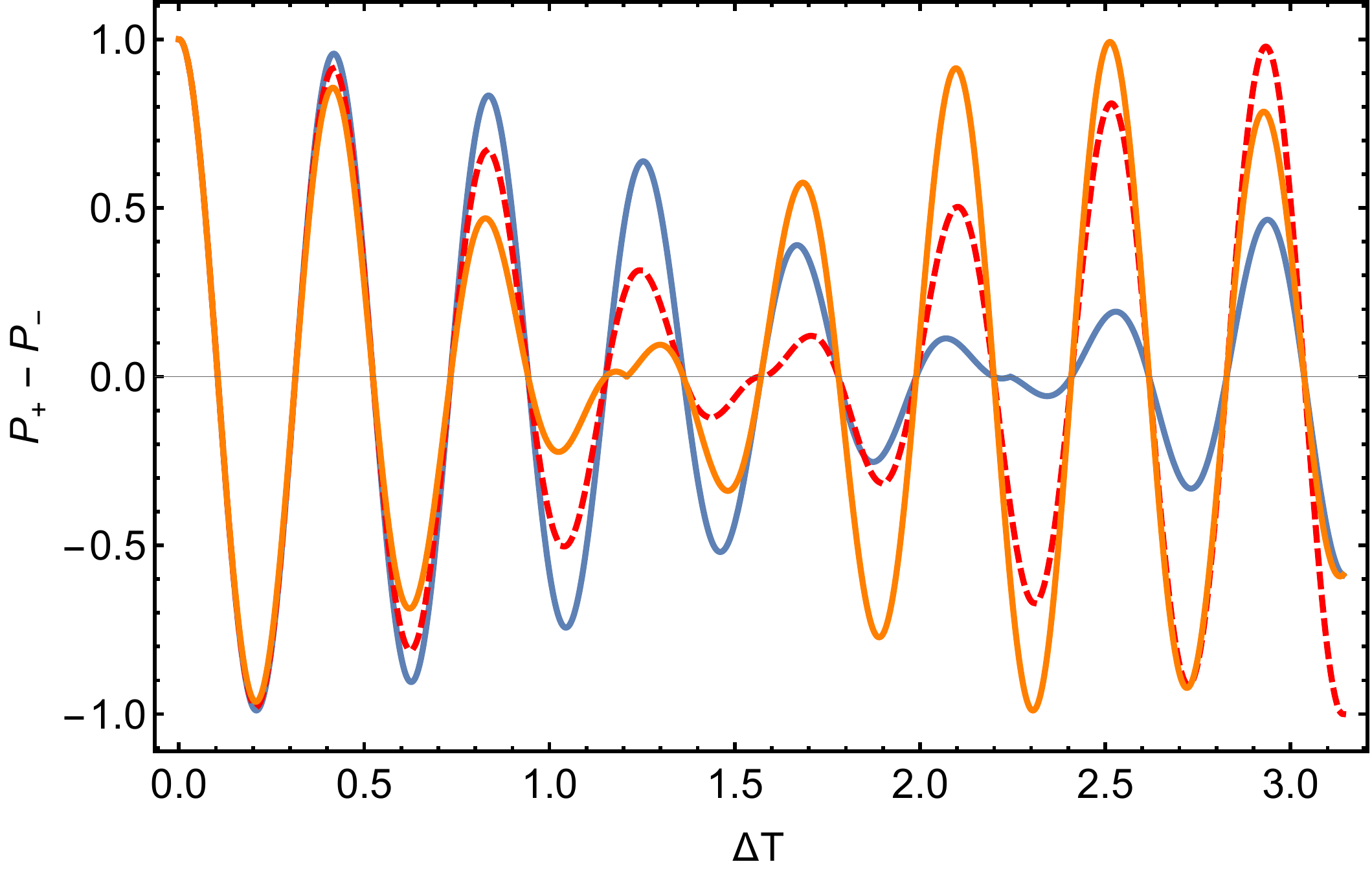}
	\centering
	\protect\caption{Behavior of the detection probabilities~\eqref{Probab-difference} in Einstein's GR (dashed red line), $\mathcal{R}^2$-gravity (orange solid line) and analytic nonlocal gravity (blue solid line). Since we are	only interested in the qualitative behavior of the decoherence, in the plot we have magnified the differences between the three cases under examination; for this reason we have rescaled $\Delta T$ so as to render it dimensionless. Moreover, to ease the comparison we have chosen the same phase for all the models. Thus, we can clearly see that for $\mathcal{R}^2$-gravity the decoherence process is faster as compared to GR, while in analytic nonlocal gravity it is slower.}\label{fig5}
\end{figure}

So far, we have only discussed the non-relativistic regime, but interesting outcomes emerge especially when relativistic effects are accounted for. Although an experimental verification of any effect beyond GR would require a non-trivial improvement of the current experimental status, we can still make a qualitative prediction. Specifically, we can understand how the decoherence associated with time dilation is dramatically influenced by the modification of the Newtonian potential.
As a matter of fact, the phenomenon of decoherence significantly depends on the strength of gravity: the weaker (stronger) the gravitational field interacts, the slower (faster) the loss of coherence becomes. In Sec.~\ref{quant-int-beyond-sec}, we have introduced both these types of theories, namely theories in which gravity turns out to be weaker at short distances with respect to GR and theories in which gravity becomes increasingly stronger. Hence, it is now clear that the loss of coherence heavily depends on the specific theory with which to describe the gravitational interaction. 

To better illustrate this concept qualitatively, we consider $\mathcal{R}^2$-gravity with the gravitational potential~\eqref{fr} and the analytic nonlocal gravity with the related metric potential~\eqref{ghost-free-pot}. It is easy to understand that, for the former model, the gravitational interaction becomes stronger at smaller scales, whereas for the latter it becomes weaker. Thus, we would expect that in $\mathcal{R}^2$-gravity the decoherence effect goes faster, whilst in analytic nonlocal gravity it goes slower. 
In Fig.~\ref{fig5}, we show the magnified behavior of the detection probability~\eqref{Probab-difference} for these two theories in comparison with the GR case. Since we are interested in comparing the interferometric visibility for several gravitational theories, we assume that some controllable phase shift is introduced, so that no difference in the oscillatory phase appears between GR and the various alternatives.

It is indeed evident that the presence of an extra attractive contribution tends to render the decoherence effect faster, while nonlocality tends to weaken gravity and slow down the loss of coherence.
This analysis might be further substantiated in the near future if the experimental status undergoes a non-trivial improvement. In fact, if the gravitational decoherence process is verified in a laboratory test, then one could experimentally discriminate among several extended theories of gravity, and set the stage for a brand-new series of experimental tests to probe new physics beyond GR.

\section{Concluding remarks \& outlook}\label{concl-sec}

In this paper, we have studied the phenomenon of quantum interference in an external gravitational field beyond Einstein's GR. After the description of our theoretical framework and the experimental apparatus, we have determined the relevant physical quantities which are measured in these laboratory tests, among which we have seen the detection probabilities and the interferometric visibility.
Subsequently, we have computed these quantities in the presence of several weak gravitational fields related to distinct extended theories of gravity. 

We have discussed both non-relativistic (COW effect) and relativistic (decoherence) implications, with an equal emphasis on both quantitative and qualitative features. 
We have noticed that, by working in the setup of the COW laboratory experiment, we can get constraints that are of the same order of the ones coming from Gravity Probe B, which is the best satellite experiment conceived so far. We have also pointed out that there are promising signals towards significant experimental improvements in the near future, which could hopefully allow to exceed the performance of torsion-balance laboratory experiments and thus give stronger constraints on departures from GR.
In particular, we have observed that, when relativistic effects are non-negligible, deviations from GR could be probed by looking at the rate of the loss of coherence induced by time dilation effects. Indeed, each theory of gravity predicts a different time scale over which the decoherence occurs, and this could be deemed as an unambiguous signature to discriminate among several extended gravitational models in a laboratory, thereby providing a unique test bench to challenge Einstein's theory.

It is worth mentioning that the proper-time difference in a Mach-Zender interferometer involving light pulses was proven to vanish in the case of a linear gravitational potential, so that the loss of coherence is only visible at non-linear order in the potential~\cite{Loriani:2019zic}. The same feature does not seem to become manifest in the case of neutron interferometry, but more detailed and quantitative future studied are surely needed.

Before concluding, let us remark that our study may potentially unravel a new and unexplored path to test and constrain new physics beyond Einstein's GR by resorting to the physics of quantum interference. In this respect, it is worth mentioning that recently Bell-type experiments were also regarded as a novel and fertile ground where to examine gravitational physics~\cite{Bittencourt:2020lgu}.

\acknowledgments
L.B. acknowledges support from JSPS and KAKENHI Grant-in-Aid for Scientific Research No.~JP19F19324.


\appendix

\section{Schr\"odinger equation in an external gravitational field}\label{schr-curved-sec}

In this Appendix, we derive the Schr\"odinger equation for a quantum system in an external gravitational field and the ensuing Hamiltonian. In particular, we compute the important equations~\eqref{schr} and~\eqref{Hamiltonian} that were used in the main text.
Let us recall that, by working with the linearized spacetime metric~\eqref{metric}, we can expand the curved d'Alembertian~\eqref{d'alembert-scalars} and write the Klein-Gordon equation~\eqref{klein-g-curved} up to the linear order in Newton's constant, thus obtaining Eq.~\eqref{klein-g-curved-lin}. We closely follow the procedure displayed in Refs.~\cite{Kiefer:1990pt,Lammerzahl:1995zz} and generalize it to the case of two distinct metric potentials $\Phi$ and $\Psi.$ 

As a first step, we cast the scalar field in a WKB-like form
\begin{equation}
\varphi(x)=e^{\frac{i}{\hbar}\tilde{\varphi}(x)}\,,\label{WKB}
\end{equation}
and consider a relativistic expansion
\begin{equation}
\tilde{\varphi}(x)=c^2\tilde{\varphi}_0(x)+\tilde{\varphi}_1(x)+\frac{1}{c^2}\tilde{\varphi}_2(x)\,,\label{relativistic-expans}
\end{equation}
where $\tilde{\varphi}_0(x)$ is taken as a real function. We can now solve the Klein-Gordon equation perturbatively in powers of the speed of light $c.$

At the order $c^4,$ the only contribution to the Klein-Gordon equation~\eqref{klein-g-curved-lin} is
\begin{equation}
\left(\vec{\nabla}\tilde{\varphi}_0\right)^2=0\quad \Rightarrow\quad \tilde{\varphi}_0\equiv \tilde{\varphi}_0(t)\,.
\end{equation}
At the order $c^2,$ we have
\begin{equation}
\left(\partial_t\tilde{\varphi}_0\right)^2-m^2=0\quad \Rightarrow \quad \tilde{\varphi}_0(t)=\pm mt+{\rm const.}\,,
\end{equation}
and we select the "minus" sign which corresponds to particles at rest with positive energy. As a consequence, up to the order $c^2$ the scalar field reads
\begin{equation}
\varphi(x)\simeq e^{-i\frac{m\,c^2}{\hbar}t}\,.
\end{equation}
At the order $c^0,$ we see that
\begin{equation}
\partial_t\tilde{\varphi}_1= -m\Phi+\frac{i\hbar}{2m}\nabla^2\Psi-\frac{1}{2m}\left(\vec{\nabla}\tilde{\varphi}_1\right)^2\,.
\end{equation}
By defining $\varphi_1=e^{\frac{i}{\hbar}\tilde{\varphi}_1}$ we can rewrite the previous equation as 
\begin{equation}
i\hbar \partial_t \varphi_1=-\frac{\hbar^2}{2m}\nabla^2\varphi_1+m\Phi\varphi_1\,,
\end{equation}
which is the non-relativistic Schr\"odinger equation for a quantum field $\varphi_1$ in an external gravitational field $\Phi.$

At the next order, relativistic contributions to the Schr\"odinger equation begin to appear. Indeed, at $c^{-2}$ the Klein-Gordon equation~\eqref{klein-g-curved-lin} gives
\begin{eqnarray}
&& \!\!\!\!-2m\partial_t\tilde{\varphi}_2 -i\hbar \partial_t^2\tilde{\varphi}_1+ (\partial_t\tilde{\varphi}_1)^2+4m\Phi \partial_t\tilde{\varphi}_1+i\hbar\nabla^2\tilde{\varphi}_2\nonumber \\[2mm]
&&\qquad-2\vec{\nabla}\tilde{\varphi}_2\cdot \vec{\nabla}\tilde{\varphi}_1 +2\Psi[i\hbar \nabla^2\tilde{\varphi}_1-(\vec{\nabla}\tilde{\varphi}_1)^2]\nonumber\\[2mm]
&&\qquad\qquad\qquad+i\hbar \vec{\nabla}(\Phi-\Psi)\cdot \vec{\nabla}\tilde{\varphi}_1=0\,.
\end{eqnarray}
By recalling the definition of $\varphi_1$ and denoting $\psi = \varphi_1 e^{\frac{i}{\hbar c^2}\tilde{\varphi}_2},$ after some algebra we can cast the previous equation in terms of the function $\psi,$ that is
\begin{eqnarray}
\!\!i\hbar \partial_t \psi(t,\vec{x})\!\!&=&\!\! \left[-\frac{\hbar^2}{2m}\nabla^2-\frac{\hbar^4}{8m^3c^2}\nabla^4+m\Phi+\frac{\hbar^2}{4mc^2}\nabla^2\Phi \right.\,\,\,\,\nonumber\\[2mm]
&-&\left.\frac{\hbar^2}{mc^2}\left(\frac{\Phi}{2}+\Psi\right)\nabla^2+\frac{\hbar^2}{2mc^2}\vec{\nabla}\Psi\cdot \vec{\nabla}   \right]\psi(t,\vec{x})\nonumber\\[2mm]
&\equiv&H \psi(t,\vec{x})\,,
\end{eqnarray}
where we have neglected orders higher than $c^{-2}$ and we have defined the Hamiltonian of a relativistic quantum system in an external gravitational field
\begin{eqnarray}
H&=& -\frac{\hbar^2}{2m}\nabla^2-\frac{\hbar^4}{8m^3c^2}\nabla^4+m\Phi+\frac{\hbar^2}{4mc^2}\nabla^2\Phi\nonumber\\[2mm]
&&-\frac{\hbar^2}{mc^2}\left(\frac{\Phi}{2}+\Psi\right)\nabla^2+\frac{\hbar^2}{2mc^2}\vec{\nabla}\Psi\cdot \vec{\nabla}  \,.
\end{eqnarray}
By using the momentum representation $\vec{p}=-i\hbar \vec{\nabla},$ we can check that the Hamiltonian becomes
\begin{eqnarray}
\!\!H\!&\!=\!& m_rc^2+\frac{p^2}{2m_r}-\frac{p^4}{8m_r^3 c^2}+m_r\Phi+\frac{1}{m_rc^2}\left(\frac{\Phi}{2}+\Psi\right)p^2\nonumber\\[2mm]
&& -\frac{1}{4m_rc^2}[p^2\Phi]-\frac{1}{2m_rc^2}[\vec{p}\,\Psi]\cdot\vec{p} \,,
\end{eqnarray}
which is precisely the one employed in the main text~\eqref{Hamiltonian}.

\section{Linearized quadratic theories of gravity}\label{gen-poisson-sec}

In what follows, we briefly review several properties of quadratic theories of gravity in the linearized regime and provide a generic integral expression for the modified Newtonian potentials.

Since we are interested in the weak-field approximation, we can expand the action~\eqref{quad-action} around Minkowski
\begin{equation}
g_{\mu\nu}=\eta_{\mu\nu}+\kappa h_{\mu\nu}\,,\label{lin-metric}
\end{equation}
where we have introduced the small metric perturbation $h_{\mu\nu}.$ In so doing, we obtain~\cite{Biswas:2011ar}
\begin{eqnarray}
\nonumber
&&\!\!\!\!\!\!\!\!\!S\!=\!\frac{1}{4}\!\int\!\! d^4x\!\left[ \frac{1}{2}h_{\mu\nu}f(\Box)\Box h^{\mu\nu}\!\!-\!h_{\mu}^{\sigma}f(\Box)\partial_{\sigma}\partial_{\nu}h^{\mu\nu}\!\!-\!\frac{1}{2}h\,g(\Box)\Box h\right.\\[4mm]
&&\!\!\!\!\!\!\!\!\!\!\!\left.+\!\,h\,g(\Box)\partial_{\mu}\partial_{\nu}h^{\mu\nu}\!+\!\frac{1}{2}h^{\lambda\sigma}\frac{f(\Box)\!-\!g(\Box)}{\Box}\partial_{\lambda}\partial_{\sigma}\partial_{\mu}\partial_{\nu}h^{\mu\nu}\right] ,
\label{lin-quad-action}
\end{eqnarray}
where $h\equiv\eta_{\mu\nu}h^{\mu\nu}$ and 
\begin{eqnarray}
f(\Box)&=&1+\frac{1}{2}F_2(\Box)\Box\,,\\[2mm]
g(\Box)&=&1-2F_1(\Box)\Box-\frac{1}{2}F_2(\Box)\Box\,.
\end{eqnarray}
By varying the action in Eq.~\eqref{lin-quad-action}, we get the field equations
\begin{eqnarray}
\nonumber
&&f(\Box)\left(\Box h_{\mu\nu}-\partial_{\sigma}\partial_{\nu}h_{\mu}^{\sigma}-\partial_{\sigma}\partial_{\mu}h_{\nu}^{\sigma}\right)\\[2mm]
\nonumber
&&+\,g(\Box)\left(\eta_{\mu\nu}\partial_{\rho}\partial_{\sigma}h^{\rho\sigma}+\partial_{\mu}\partial_{\nu}h-\eta_{\mu\nu}\Box h\right)\\[2mm]
&&+\, \frac{f(\Box)-g(\Box)}{\Box}\partial_{\mu}\partial_{\nu}\partial_{\rho}\partial_{\sigma}h^{\rho\sigma}=-2\kappa^2 T_{\mu\nu},
\label{lin-field-eq}
\end{eqnarray}
with $T_{\mu\nu}$ being the stress-energy tensor for the matter action $S_{\rm m}$
\begin{equation}
T_{\mu\nu}=-\frac{2}{\sqrt{-g}}\frac{\delta S_{\rm m}}{\delta g^{\mu\nu}}\simeq 2\frac{\delta S_{\rm m}}{\delta h^{\mu\nu}}\,.
\end{equation}
For our purposes, it suffices to consider a configuration in which the source can be well-approximated by a static and pressure-less point-like object. The stress-energy tensor associated with such a source is
\begin{equation}
T_{\mu\nu}=Mc^2\delta_{\mu}^0\delta_{\nu}^0\delta^{(3)}(\vec{r}).
\end{equation}
In light of this choice, by resorting to the generic shape for the line element given in Eq.~\eqref{metric}, we have $\kappa h_{00}=2\Phi/c^2$ and $\kappa h_{ij}=2\Psi\delta_{ij}/c^2,$ and hence the modified Poisson equations for the two metric potentials read
\begin{eqnarray}
\label{potpot}
\!\!\!\!\!\frac{f(\nabla^2)[f(\nabla^2)-3g(\nabla^2)]}{f(\nabla^2)-2g(\nabla^2)}\nabla^2\Phi(r)&=&8\pi GM\delta^{(3)}(\vec{r})\,,\quad\,\,\,\\[2mm]
\!\!\!\!\!\frac{f(\nabla^2)[f(\nabla^2)-3g(\nabla^2)]}{g(\nabla^2)}\nabla^2\Psi(r)&=&- 8\pi GM\delta^{(3)}(\vec{r})\,,\quad\,\,\,
\label{field-eq-pot}
\end{eqnarray}
where we assume $\Box\simeq \nabla^2$ due to the staticity requirement.

The differential equations~\eqref{potpot} and~\eqref{field-eq-pot} can be solved  by using the Fourier transform method. Indeed, by using spherical coordinates we can exhibit the solutions for the metric potentials in the following integral form:
\begin{equation}
\begin{array}{rl}
\Phi(r)
\label{pot1}
=& \displaystyle-\frac{4GM}{\pi r}\int_0^{\infty}dk\frac{f-2g}{f(f-3g)}\frac{{\rm sin}(kr)}{k}\,,\\[3mm]
\Psi(r)=& \displaystyle\frac{4GM}{\pi r}\int_0^{\infty}dk\frac{g}{f(f-3g)}\frac{{\rm sin}(kr)}{k}\,,
\end{array}
\end{equation}
where $f=f(-k^2)$ and $g=g(-k^2)$ are functions of the Fourier momentum squared  $k^2.$ These last two formulas were exploited in Sec.~\ref{quant-int-beyond-sec} to compute the modified gravitational potentials in each extended theory of gravity. As a consistency check, note that, as $f=g=1$ ($F_1=F_2=0$), we recover the Newtonian potential $\Phi=\Psi=-GM/r.$


\end{document}